\documentclass[a4paper,12pt]{article}
\usepackage{ascmac}
\usepackage{color}
\usepackage{multirow}
\newcommand{\BQED}{\hfill \hbox{\rule{8pt}{8pt}}}

\newenvironment{namelist}[1]{%
\begin{list}{}
  { 
	\settowidth{\labelwidth}{#1}
	\setlength{\leftmargin}{1.1\labelwidth}}
  \setlength{\itemsep}{0cm}
}{% 
\end{list}}
\catcode`\@=11
\def\newexample#1{\@ifnextchar[{\@oexm{#1}}{\@nexm{#1}}}

\def\@nexm#1#2{%
\@ifnextchar[{\@xnexm{#1}{#2}}{\@ynexm{#1}{#2}}}

\def\@xnexm#1#2[#3]{\expandafter\@ifdefinable\csname #1\endcsname
{\@definecounter{#1}\@addtoreset{#1}{#3}%
\expandafter\xdef\csname the#1\endcsname{\expandafter\noexpand
  \csname the#3\endcsname \@exmcountersep \@exmcounter{#1}}%
\global\@namedef{#1}{\@exm{#1}{#2}}\global\@namedef{end#1}{\@endexample}}}

\def\@ynexm#1#2{\expandafter\@ifdefinable\csname #1\endcsname
{\@definecounter{#1}%
\expandafter\xdef\csname the#1\endcsname{\@exmcounter{#1}}%
\global\@namedef{#1}{\@exm{#1}{#2}}\global\@namedef{end#1}{\@endexample}}}

\def\@oexm#1[#2]#3{\expandafter\@ifdefinable\csname #1\endcsname
  {\global\@namedef{the#1}{\@nameuse{the#2}}%
\global\@namedef{#1}{\@exm{#2}{#3}}%
\global\@namedef{end#1}{\@endexample}}}

\def\@exm#1#2{\refstepcounter
    {#1}\@ifnextchar[{\@yexm{#1}{#2}}{\@xexm{#1}{#2}}}

\def\@xexm#1#2{\@beginexample{#2}{\csname the#1\endcsname}\ignorespaces}
\def\@yexm#1#2[#3]{\@opargbeginexample{#2}{\csname
       the#1\endcsname}{#3}\ignorespaces}

\def\@exmcounter#1{\noexpand\arabic{#1}}
\def\@exmcountersep{.}
\def\@beginexample#1#2{\trivlist \item[\hskip 
\labelsep{\bf #1\ #2:}]}
\def\@opargbeginexample#1#2#3{\trivlist
      \item[\hskip \labelsep{\bf #1\ #2\ }#3{\bf :}]}
\def\@endexample{\endtrivlist}

\catcode`\@=12

\newtheorem{lemma}{{\bf Lemma}}[section]
\newtheorem{thm}{{\bf Theorem}}[section]
\newtheorem{df}{{\bf Definition}}[section]
\newtheorem{cor}{{\bf Corollary}}[section]
\newtheorem{prop}{{\bf Proposition}}[section]

\newtheorem{claim}{{\bf Claim}}[section]

\newcommand{\msc}[1]{\mbox{{\sc #1}}}
\newcommand{\leng}[1]{|{#1}|}
\newcommand{\bm}[1]{\mbox{\boldmath{$#1$}}}
\catcode`\@=11

\def\@xnthm#1#2[#3]{\expandafter\@ifdefinable\csname #1\endcsname
{\@definecounter{#1}\@addtoreset{#1}{#3}%
\expandafter\xdef\csname the#1\endcsname{\expandafter\noexpand
\bf \csname the#3\endcsname \@thmcountersep \@thmcounter{#1}}%
\global\@namedef{#1}{\@thm{#1}{#2}}\global\@namedef{end#1}{\@endtheorem}}}

\def\@ynthm#1#2{\expandafter\@ifdefinable\csname #1\endcsname
{\@definecounter{#1}%
\expandafter\xdef\csname the#1\endcsname{\bf \@thmcounter{#1}}%
\global\@namedef{#1}{\@thm{#1}{#2}}\global\@namedef{end#1}{\@endtheorem}}}

\def\@begintheorem#1#2{\trivlist \item[\hskip 
\labelsep{\bf #1~#2:}]\sl}
\def\@opargbegintheorem#1#2#3{\trivlist
      \item[\hskip \labelsep{\bf #1~#2}~#3{\bf :}]\sl}

\catcode`\@=12

\newbox\rubisita
\newbox\rubiue
\newdimen\rubiw
\def\rubi#1#2{{\setbox\rubisita=\hbox{#1}\setbox\rubiue=\hbox{\tiny #2}%
\ifdim \wd\rubisita>\wd\rubiue\rubiw=\wd\rubisita\else\rubiw=\wd\rubiue\fi%
\kanjiskip=0pt plus1fil%
\setbox\rubisita=\hbox to \rubiw{\hfil#1\hfil}%
\setbox\rubiue=\hbox to \rubiw{\tiny\hfil#2\hfil}%
\vbox{\offinterlineskip\box\rubiue\break\box\rubisita}}}
\begin{document}
%
%\vspace*{-1.7cm}
%
%\begin{flushright}
%
%October 4, 2017
%
%\end{flushright}\smallskip
%
\begin{center}
{\Large {\bf On the Hardness of Deriving the 
Arithmetic Mean Component Competitive Ratio}}\bigskip\\
%
%\begin{tabular}{ccc}
%
{\sc Toshiya Itoh}\footnote[1]{~
Dept. of Mathematical and Computing Science, 
Tokyo Institute of Technology, Japan.} \hspace*{1.0cm}
{\sc Yoshinori Takei}\footnote[2]{~
Dept. of Electrical and Information Engineering,
National Institute of Technology, Akita College,  Japan.}\\
{\sf titoh@c.titech.ac.jp} \hspace*{0.75cm} {\sf ytakei@akita-nct.ac.jp}
%
%\end{tabular}
%
\end{center}\medskip
{\sf Abstract:} For the multi-objective time series search problem, 
Hasegawa and Itoh [Theoretical Computer Science, Vo.718, pp.58-66, 2018] 
presented the best possible online algorithm~{\it balanced price policy\/} 
({\sc bpp}) 
for any monotone function $f: {\bf R}^{k} \to {\bf R}$.  
Specifically, the competitive~ratio with respect to the monotone function 
$f(c_{1},\ldots,c_{k})=(c_{1}+\cdots+c_{k})/k$ is referred to as the arithmetic 
mean component competitive ratio.  
Hasegawa and Itoh derived the closed formula~of~the arithmetic mean component 
competitive ratio for $k=2$, but it has not been known~for~any~integer $k \geq 3$. In this 
paper, we show that it is ${\cal NP}$-hard to derive closed formulas of the arithmetic 
mean component competitive ratio for general integer $k\geq 2$.  
On the the hand, we 
derive closed formulas of the 
arithmetic mean component competitive ratio for $k=3$  and $k=4$. \medskip
\section{Introduction} \label{sec-introduction}
%==============================================
%
There are many single-objective online optimization problems such as  
paging and caching~(see \cite{Y}~for a survey), 
metric task systems (see \cite{K} for a survey), 
asset conversion problems (see \cite{MAS} for a survey), 
buffer management of network switches (see \cite{G} for a survey), etc., and 
Sleator and Tarjan \cite{ST} introduced
a notion of competitive analysis 
%are fundamental  in 
%computing, communicating, and many other practical systems. 
to measure the efficiency of online algorithms for single-objective 
online optimization problems. 
%Since then extensive research has been made for diverse areas 
%of single-objective online optimization problems, e.g.,  
For online problems of multi-objective nature, 
%however, general framework of 
%competitive analysis and definition of competitive ratio for 
%multi-objective online problems are not known. 
Tiedemann, et al. \cite{TIS} presented a framework of 
multi-objective online problems as an online version of multi-objective 
optimization problems \cite{E} and formulated 
a notion of the competitive ratio for multi-objective online problems 
as the extension of the competitive ratio for single-objective online problems. 
On defining the competitive ratio for %muti-objective 
$k$-objective online problems, Tiedemann, et al. \cite{TIS} regarded multi-objective online 
problems as a family of (possibly dependent) 
single-objective online problems and applied a monotone 
function $f:{\bf R}^{k}\to {\bf R}$ 
to the family of the single-objective online problems. Let 
${\cal A}$ be an algorithm for a %multi-objective 
$k$-objective online 
problem. Then we regard the algorithm ${\cal A}$ as a family of algorithms 
${\cal A}_{i}$ for the $i$th objective. 
For $c_{1},\ldots,c_{k}$, where $c_{i}$ is the competitive ratio of 
the algorithm ${\cal A}_{i}$,  
we~say~that~the~algorithm~${\cal A}$ is $f(c_{1},\ldots,c_{k})$-competitive 
with respect to a monotone function $f: {\bf R}^{k} \to {\bf R}$. 
In fact, Tiedemann, et al. \cite{TIS} defined the competitive ratio by several 
monotone~(continuous) functions,~e.g.,~the worst component competitive ratio by $f_{1}(c_{1},\ldots,c_{k}) = 
\max(c_{1},\ldots,c_{k})$,~the~arithmetic mean component competitive ratio by 
$f_{2}(c_{1},\ldots,c_{k})= (c_{1}+\cdots+c_{k})/k$, and 
the geometric mean component competitive ratio by 
$f_{3}(c_{1},\ldots,c_{k}) = (c_{1}\times \cdots \times c_{k})^{1/k}$. 
%
%=====================================================
\subsection{Previous Work} \label{subsec-previous}
%=====================================================
%
%=========================================================
%\subsection{Multi-Objective Time Series Search Problem}
%\label{subsec-time-series}
%=========================================================
%
A single-objective time series search problem (initially investigated 
by El-Yaniv, et al. \cite{Eetal}) is defined as follows: 
Let $\msc{alg}$ be a player that searches for the maximum price in a sequence of prices. 
At the beginning of each time period $t \in \{1,\ldots,T\}$, the player 
$\msc{alg}$~receives~a~price~$p_{t} \in {\bf R}_{+}$ and must decide 
whether to accept or reject the price $p_{t}$. 
Assume that prices are chosen from the interval $\msc{itv}=[m,M]$, where 
$0<m \leq M$. %, and that $m$ and $M$ are known to the player 
%$\msc{alg}$\footnote{~We can show that if only the fluctuation ratio 
%$\phi=M/m$ is known (but not $m$ or $M$) to the player {\sc alg}, then no better 
%competitive  ratio than the trivial one of $\phi$ is achievable.}. 
If the player $\msc{alg}$ accepts $p_{t}$, then 
the game~ends and the return for $\msc{alg}$ is $p_{t}$. 
If the player $\msc{alg}$ rejects $p_{t}$ for every $t \in \{1,\ldots,T\}$, 
then~the~return~for $\msc{alg}$ is defined to be $m$. Let $r=M/m$ be the {\it fluctuation ratio\/} 
of possible~prices.~For~the~case that 
$m$ and $M$ are known to online algorithms, 
El-Yaniv, et al. \cite{Eetal} presented a (best possible) 
deterministic algorithm {\it reservation price policy\/} {\sc rpp}, 
which is $\sqrt{r}$-competitive,~and~a~randomi\-zed algorithm {\it exponential threshold\/} 
{\sc expo}, which~is~$O(\log r)$-competitive\footnote[3]{~We can show that if only the fluctuation ratio 
$r=M/m$ is known (but not $m$ or $M$) to the (deterministic) player {\sc alg}, then no better 
competitive  ratio than the trivial one of $r$ is achievable.}. 

As a natural extension of the 
single-objective time series search problem, 
a multi-objective ($k$-objective) time series search 
problem \cite{TIS} can be defined as follows: 
At the beginning of each time period $t \in \{1,\ldots,T\}$, 
the player $\msc{alg}_{k}$ receives a price vector 
$\vec{p}_{t}=(p_{t}^{1},\ldots,p_{t}^{k}) \in {\bf R}_{+}^{k}$~and must 
decide whether to accept or reject the price vector $\vec{p}_{t}$. 
As in the case of a single-objective time series search problem, 
assume that price $p_{t}^{i}$ is chosen from the interval 
$\msc{itv}_{i}=[m_{i},M_{i}]$, where $0 < m_{i} \leq M_{i}$ for each $i \in \{1,\ldots,k\}$, 
and that the player $\msc{alg}_{k}$ knows $m_{i}$ and $M_{i}$~for~each $i \in \{1,\ldots,k\}$. 
If the player $\msc{alg}_{k}$ accepts $\vec{p}_{t}$, 
then the game ends and the return~for~$\msc{alg}_{k}$~is~$\vec{p}_{t}$. 
If the player $\msc{alg}_{k}$ rejects $\vec{p}_{t}$ for every $t \in \{1,\ldots,T\}$, then 
the return for $\msc{alg}_{k}$ is defined~to~be~the {\it minimum\/} price 
vector $\vec{p}_{\rm min}=(m_{1},\ldots,m_{k})$. 
Without loss of generality, assume that~$M_{1}/m_{1}\geq \cdots \geq M_{k}/m_{k}$. 
For the 
case that all of $\msc{itv}_{1}=[m_{1},M_{1}],\ldots,
\msc{itv}_{k}=[m_{k},M_{k}]$ are {\it real\/} intervals, 
Tiedemann, et al. \cite{TIS} presented best possible 
online algorithms for the multi-objective time series search problem with 
respect to the monotone functions 
$f_{1}$, $f_{2}$, and $f_{3}$, i.e., 
the best possible online algorithm {\sc rpp-high} for the multi-objective time series search 
problem with respect to the monotone function $f_{1}$ 
\cite[Theorems 1 and 2]{TIS}, 
the best possible online algorithm {\sc rpp-mult} for the bi-objective time series search 
problem with respect to the monotone function~$f_{2}$~\cite[Theorems 3 and 4]{TIS}
and the best possible online algorithm {\sc rpp-mult} for the bi-objective time series search 
problem with respect to the monotone function $f_{3}$ \cite[\S3.2]{TIS}.
Recently, Hasegawa and Itoh \cite{HI} 
presented the deterministic online algorithm {\it balanced price policy\/} {\sc bpp} and 
showed that {\sc bpp} is best possible for any monotone function $f: {\bf R}^{k}\to {\bf R}$ 
and for any integer $k \geq 2$. 
%
%===========================================================
\subsection{Our Contribution} \label{subsec-contribution}
%===========================================================
%
In~this~paper,~we~show that it is ${\cal NP}$-hard to derive closed formulas of  
the arithmetic mean~component~competitive ratio for general integer $k\geq 2$, but 
this does not necessarily imply that~it~is difficult %${\cal NP}$-hard 
to write out closed formulas of the 
arithmetic mean component competitive ratio for a fixed integer $k\geq 2$. 
%
%this does not necessarily imply that~it~is ${\cal NP}$-hard to derive a closed formula of  
%the arithmetic mean~component~competitive ratio for a fixed integer $k\geq 2$. 
In fact, 
Hasegawa and Itoh \cite{HI} derived a closed formula of the arithmetic mean component 
competitive ratio for $k=2$. In this paper, we also derive 
closed formulas of 
the arithmetic mean component competitive ratio for 
$k=3$  and $k=4$. 
%
%==============================================
\section{Preliminaries} \label{sec-preliminary}
%==============================================
%
%==============================================
\subsection{Notations} \label{subsec-notation}
%==============================================
%
%For any pair of integers $a\leq b$, let $[a,b]=\{a,\ldots,b\}$. 
Let $k \geq 2$ be an integer. For each $1 \leq i  \leq k$, let 
$\msc{itv}_{i}=[m_{i},M_{i}]$ be the interval~of~the~$i$th component 
of price vectors for the $k$-objective time series search problem, and we use 
$r_{i} = M_{i}/m_{i}$~to denote the {\it fluctuation ratio\/} of the interval $\msc{itv}_{i}=[m_{i},M_{i}]$. 
Without loss of generality,~we~assume that $r_{1}\geq \cdots \geq r_{k}\geq 1$. 
%, i.e., $M_{1}/m_{1} \geq \cdots \geq M_{k}/m_{k}\geq 1$. 
%
For any monotone continuous function $f: {\bf R}^{k}\to {\bf R}$, 
define 
\[
{\cal S}_{f}^{k}=\left\{(x_{1},\ldots,x_{k}) \in \msc{itv}_{1}\times \cdots \times \msc{itv}_{k}: 
f \left(\frac{x_{1}}{m_{1}},\ldots,\frac{x_{k}}{m_{k}}\right)= 
f\left(\frac{M_{1}}{x_{1}},\ldots,\frac{M_{k}}{x_{k}}\right)\right\},
\]
and Hasegawa and Itoh \cite[Theorem 3.1]{HI} 
showed that with respect to any monotone %(continuous) 
function~$f: {\bf R}^{k}\to {\bf R}$, the 
competitive ratio for the $k$-objective time series search problem is given~by 
\[
z_{f}^{k}=\sup_{(x_{1},\ldots,x_{k}) \in {\cal S}_{f}^{k}} 
f\left(\frac{M_{1}}{x_{1}},\ldots,\frac{M_{k}}{x_{k}}\right).
\]
For the  
monotone function $f(c_{1},\ldots,c_{k})=(c_{1}+\cdots+c_{k})/k$, 
we refer to $z_{f}^{k}$ as the 
arithmetic~mean component competitive ratio for the $k$-objective time series search problem. 
In the rest of this paper, we focus on the function 
$f(c_{1},\ldots,c_{k})=(c_{1}+\cdots+c_{k})/k$. 
For the bi-objective time~series search problem, 
Tiedemann, et al.  \cite[Theorem 3]{TIS} %formulated multi-objective online problems and 
derived the arithmetic mean~component competitive ratio 
$(r_{1}r_{2})^{1/4}$, which is disproved by 
Hasegawa and Itoh \cite[Theorem 4.1]{HI},~i.e., 
\begin{equation}
z_{f}^{2}=\frac{1}{4}\cdot \left\{\sqrt{
4 r_{1}+(r_{2}-1)^{2}}+ (r_{2}+1)\right\} \geq  
(r_{1}r_{2})^{1/4}, \label{eq-k=2}
\end{equation}
where the equality holds in the last inequality of Eq.(\ref{eq-k=2}) if $r_{1}=r_{2}=1$. 

In this paper, we show that it is ${\cal NP}$-hard~to derive closed formulas of  
the arithmetic mean component competitive ratio for general integer $k\geq 2$. 
%
%In this paper, we will derive closed formulas of the 
%arithmetic mean component competitive~ratio for 
%the $k$-objective time series search problem for an 
%integer $k =3$ (see Theorems \ref{thm-k=3(1)} and~\ref{thm-k=3(2)}~in Section 
%\ref{sec-k=3}) and for an integer $k=4$ (see Theorems 
%\ref{thm-k=4(1)}, \ref{thm-k=4(2)},  and \ref{thm-k=4(3)} in Section 
%\ref{sec-k=4}). 
%
For %the function 
$f(c_{1},\ldots,c_{k})=(c_{1}+\cdots+c_{k})/k$,~it~is immediate to see that 
${\cal S}_{f}^{k}$ and $z_{f}^{k}$ are given~by  
\begin{eqnarray*}
{\cal S}_{f}^{k} & = & \left\{(x_{1},\ldots,x_{k}) \in\msc{itv}_{1}\times \cdots \times \msc{itv}_{k}: 
\frac{1}{k}\left(\frac{x_{1}}{m_{1}}+\cdots +\frac{x_{k}}{m_{k}}\right)= 
\frac{1}{k}\left(\frac{M_{1}}{x_{1}}+\cdots +\frac{M_{k}}{x_{k}}\right)\right\}\nonumber\\
& = & \left\{(x_{1},\ldots,x_{k}) \in\msc{itv}_{1}\times \cdots \times \msc{itv}_{k}: 
\left(\frac{x_{1}}{m_{1}}-\frac{M_{1}}{x_{1}}\right)+ \cdots +
\left(\frac{x_{k}}{m_{k}}-\frac{M_{k}}{x_{k}}\right)=0\right\};\\%\label{eq-Sf}\\
z_{f}^{k} & = & \sup_{(x_{1},\ldots,x_{k}) \in {\cal S}_{f}^{k}} 
\frac{1}{k}\left(\frac{M_{1}}{x_{1}}+\cdots+\frac{M_{k}}{x_{k}}\right)\nonumber\\
& = & \frac{1}{k}\sup_{(x_{1},\ldots,x_{k}) \in {\cal S}_{f}^{k}} 
\left(\frac{M_{1}}{x_{1}}+\cdots +\frac{M_{k}}{x_{k}}\right)\nonumber\\
& = & \frac{1}{k}\sup_{(x_{1},\ldots,x_{k}) \in {\cal S}_{f}^{k}} \left\{ \frac{1}{2}
\left(\frac{x_{1}}{m_{1}}+\cdots+\frac{x_{k}}{m_{k}}\right) + 
\frac{1}{2}
\left(\frac{M_{1}}{x_{1}}+\cdots+\frac{M_{k}}{x_{k}}\right)\right\}\nonumber\\
& = & \frac{1}{2k}\sup_{(x_{1},\ldots,x_{k}) \in {\cal S}_{f}^{k}} \left\{ 
\left(\frac{x_{1}}{m_{1}}+\frac{M_{1}}{x_{1}}\right)+ \cdots+
\left(\frac{x_{k}}{m_{k}}+\frac{M_{k}}{x_{k}}\right) \right\}, %\label{eq-zf}
\end{eqnarray*}
respectively. For each $i \in \{1,\ldots,k\}$, let $\alpha_{i} = \sqrt{M_{i}/m_{i}}=\sqrt{r_{i}}$. 
Since $r_{1}\geq \cdots \geq r_{k}\geq 1$,~we~have that 
$\alpha_{1}\geq \cdots \geq \alpha_{k}\geq 1$. For $x > 0$, let 
$\phi(x)=\frac{x-x^{-1}}{2}$~and~for~each~$i \in \{1,\ldots,k\}$, let 
\begin{equation}
\xi_{i} = \alpha_{i} \phi\left(\frac{x_{i}}{\sqrt{m_{i}M_{i}}}\right). \label{eq-correspond}
\end{equation}
Note that the function $\phi$ is monotonically increasing and it is immediate 
that $\phi(x^{-1})=-\phi(x)$. So the correspondence $x_{i}\to \xi_{i}$ in Eq.(\ref{eq-correspond}) 
bijectively maps the interval $\msc{itv}_{i}=[m_{i},M_{i}]$ to 
\[
\left[\alpha_{i} \phi\left(\frac{m_{i}}{\sqrt{m_{i}M_{i}}}\right), 
\alpha_{i} \phi\left(\frac{M_{i}}{\sqrt{m_{i}M_{i}}}\right)\right] = 
[\alpha_{i}\phi(\alpha_{i}^{-1}),\alpha_{i}\phi(\alpha_{i})] = 
[-\alpha_{i}\phi(\alpha_{i}),\alpha_{i}\phi(\alpha_{i})]. %\label{eq-itv}
\]
For simplicity, set $\beta_{i}=\alpha_{i}\phi(\alpha_{i})$. Then for each 
$1 \leq i \leq k$, we have that 
$2\beta_{i}=\alpha_{i}^{2}-1=r_{i}-1$ and let $\msc{itv}_{i}'=[-\alpha_{i}\phi(\alpha_{i}), 
\alpha_{i}\phi(\alpha_{i})]=[-\beta_{i},\beta_{i}]$. Note that 
$\beta_{1}\geq \cdots \geq \beta_{k}\geq 0$. 
%
%=====================================================
\subsection{Observations} \label{subsec-formulation}
%=====================================================
%
In this subsection, we present several observations that are crucial in the subsequent 
discussions. 
%
%\newpage
%
\begin{prop} \label{prop-correspond}
Assume that the correspondence $x_{i} \to \xi_{i}$ is given by 
{\rm Eq.(\ref{eq-correspond})}~for~each~$1 \leq i \leq k$. 
Then $(x_{1},\ldots,x_{k}) \in {\cal S}_{f}^{k}$ iff  
both of the following conditions hold$:$ {\rm (i)} $\xi_{i} \in [-\beta_{i},\beta_{i}]$ 
for each $1 \leq i \leq k$ and {\rm (ii)} $\xi_{1}+\cdots+\xi_{k}=0$. 
\end{prop}
{\bf Proof:} From the definition of the correspondence by Eq.(\ref{eq-correspond}), 
we have that 
\[
\frac{1}{2} \sum_{i=1}^{k} \left(\frac{x_{i}}{m_{i}}-\frac{M_{i}}{x_{i}}\right)=
\sum_{i=1}^{k} \sqrt{\frac{M_{i}}{m_{i}}} \cdot \frac{1}{2} \left(
\frac{x_{i}}{\sqrt{m_{i}M_{i}}}-\frac{\sqrt{m_{i}M_{i}}}{x_{i}}\right)
=\sum_{i=1}^{k}  \alpha_{i}\phi\left(\frac{x_{i}}{\sqrt{m_{i}M_{i}}}\right)
= \sum_{i=1}^{k} \xi_{i}.
\]
Then it is easy to see that 
$(x_{1},\ldots,x_{k}) \in {\cal S}_{f}^{k}$ iff 
both of the conditions (i) and (ii) hold. \BQED\medskip%
%
%\newpage

For each $1 \leq i \leq k$, let $\msc{itv}_{i}'=[-\beta_{i},\beta_{i}]$. 
Define $H(x_{1},\ldots ,x_{k})$, $G(\xi_{1},\ldots,\xi_{k})$, 
${\cal T}_{f}^{k}$ as follows: 
\begin{eqnarray*}
H(x_{1},\ldots,x_{k}) & = & \frac{1}{2k} \cdot \sum_{i=1}^{k} \left(\frac{x_{i}}{m_{i}}+\frac{M_{i}}{x_{i}}\right);\\
G(\xi_{1}\ldots,\xi_{k}) & = & \frac{1}{k} \cdot \sum_{i=1}^{k} \sqrt{\alpha_{i}^{2}+\xi_{i}^{2}}
%= \frac{2}{2k} \cdot \sum_{i=1}^{k} \sqrt{\alpha_{i}^{2}+\xi_{i}^{2}}
= \frac{1}{2k} \cdot \left(2 \sum_{i=1}^{k} \sqrt{\alpha_{i}^{2}+\xi_{i}^{2}}\right);\\
{\cal T}_{f}^{k} & = & \{(\xi_{1},\ldots,\xi_{k}) \in \msc{itv}_{1}'\times \cdots \times \msc{itv}_{k}': 
\xi_{1}+\cdots+\xi_{k}=0\}. 
\end{eqnarray*}
\begin{prop} \label{prop-max}
Assume that the correspondence $x_{i} \to \xi_{i}$ is given by 
{\rm Eq.(\ref{eq-correspond})}~for~each~$ 1 \leq i \leq k$. Then the 
problem of maximizing the function $H(x_{1},\ldots,x_{k})$ 
over ${\cal S}_{f}^{k}$ is equivalent to the problem of maximizing the function 
$G(\xi_{1},\ldots,\xi_{k})$ over ${\cal T}_{f}^{k}$. 
\end{prop}
{\bf Proof:} By straightforward calculations, we have that for each 
$1 \leq i \leq k$, 
\begin{eqnarray*}
\frac{x_{i}}{m_{i}} + \frac{M_{i}}{x_{i}}  &= &
2 \sqrt{\frac{M_{i}}{m_{i}}} \cdot
\frac{\frac{x_{i}}{\sqrt{m_{i}M_{i}}}+
\frac{\sqrt{m_{i}M_{i}}}{x_{i}}}{2} =
2\alpha_{i} 
\sqrt{%
\left(\frac{\frac{x_{i}}{\sqrt{m_{i}M_{i}}}+
\frac{\sqrt{m_{i}M_{i}}}{x_{i}}}{2} \right)^{2}}\\
& = & 2\alpha_{i} 
\sqrt{1+
\left(\frac{\frac{x_{i}}{\sqrt{m_{i}M_{i}}}-
\frac{\sqrt{m_{i}M_{i}}}{x_{i}}}{2} \right)^{2}}
= 2\alpha_{i} 
\sqrt{1+\phi^{2}
\left(
\frac{x_{i}}{\sqrt{m_{i}M_{i}}} \right)}\\
& = & 2 
\sqrt{\alpha_{i}^{2}+\alpha_{i}^{2}\phi^{2}
\left(
\frac{x_{i}}{\sqrt{m_{i}M_{i}}} \right)}= 2 \sqrt{\alpha_{i}^{2}+\xi_{i}^{2}}, 
\end{eqnarray*} 
where the last equality follows from the correspondence $x_{i}\to \xi_{i}$  in
Eq.(\ref{eq-correspond}). Thus it is immediate that 
the problem of maximizing the function $H(x_{1},\ldots,x_{k})$ 
over ${\cal S}_{f}^{k}$ is equivalent to the problem of maximizing the function 
$G(\xi_{1},\ldots,\xi_{k})$ over ${\cal T}_{f}^{k}$. \BQED\medskip%

For each $1 \leq i \leq k$, we say that $\xi_{i} \in 	\msc{itv}_{i}'$ is {\it filled\/} if 
$\xi_{i}=\beta_{i}$ or $\xi_{i}=-\beta_{i}$ 
(and~we~say~that~$\xi_{i} \in 	\msc{itv}_{i}'$  is {\it unfilled\/} if it is not filled). 
Then we can show the following lemma: 
%The following lemma plays a 
%crucial role in the subsequent discussions. 
%
\begin{lemma} \label{lemma-crucial}
If $\vec{\xi}^{*}=(\xi_{1}^{*},\ldots,\xi_{k}^{*}) \in {\cal T}_{f}^{k}$ maximizes the function 
$G(\xi_{1},\ldots,\xi_{k})$, then {\rm (i)} there exists at most~a single {\sf unfilled} variable  $\xi_{h}^{*}$
%such that $\xi_{j}$ is unfilled 
and {\rm (ii)} %$(-\xi_{1},\ldots,-\xi_{k}) \in {\cal T}_{f}^{k}$ 
$-\vec{\xi}^{*} \in {\cal T}_{f}^{k}$ maximizes the function~$G(\xi_{1},\ldots,\xi_{k})$. 
\end{lemma}
{\bf Proof:} For the statement (i), assume that there exist 
two~distinct~unfilled variables~$\xi_{j_{1}}^{*}$~and~$\xi_{j_{2}}^{*}$, i.e., 
$-\beta_{j_{1}}<\xi_{j_{1}}^{*}<\beta_{j_{1}}$ and 
$-\beta_{j_{2}}<\xi_{j_{2}}^{*}<\beta_{j_{2}}$. 
%$1 \leq j_{1}<j_{2}\leq k$. 
%such that $\xi_{j_{1}}$ and $\xi_{j_{2}}$ are unfilled. 
So we have that 
\[
G(\xi_{1}^{*},\ldots,\xi_{k}^{*})= \frac{1}{2k}\cdot 2\left(\sqrt{\alpha_{j_{1}}^{2}+(\xi_{j_{1}}^{*})^{2}} + 
\sqrt{\alpha_{j_{2}}^{2}+(\xi_{j_{2}}^{*})^{2}} +
\sum_{j \in \{1,\ldots,k\}\setminus \{j_{1},j_{2}\}} 
\sqrt{\alpha_{j}^{2}+\beta_{j}^{2}}\right).
\]
Then there exists $\eta\neq 0$ such that $\xi_{j_{1}}^{*}+\eta \in \msc{itv}_{j_{1}}'$ 
and $\xi_{j_{2}}^{*}-\eta \in \msc{itv}_{j_{2}}'$. 
Let $\xi_{j_{1}}'=\xi_{j_{1}}^{*}+\eta$,~$\xi_{j_{2}}'=\xi_{j_{2}}^{*}-\eta$, 
and $\xi_{j}'=\xi_{j}^{*}$ for each $j \in \{1,\ldots,k\}
\setminus\{j_{1},j_{2}\}$. It is immediate that 
$(\xi_{1}',\ldots,\xi_{k}')\in {\cal T}_{f}^{k}$. 
For~the rest of the proof, we use the following claim~(the proof of the claim 
is given in Appendix~\ref{append-claim}). 
\begin{claim} \label{claim-tech}
For $a,b,c,d>0$, let $h(x)=\sqrt{a^{2}+(b+x)^{2}}+\sqrt{c^{2}+(d-x)^{2}}$. Then 
$h'(x)$~and~$h''(x)$ are continuous, and 
{\rm (i)} ${\rm sgn}~h'(0)={\rm sgn} (cb-ad)$ and {\rm (ii)} $h''(0)>0$. 
\end{claim}

For Claim \ref{claim-tech}, set $a=\alpha_{j_{1}}$, $b=\xi_{j_{1}}^{*}$, 
$c=\alpha_{j_{2}}$, and $d=\xi_{j_{2}}^{*}$. If $h'(0)=
{\rm sgn}(\alpha_{j_{2}}\xi_{j_{1}}^{*}-\alpha_{j_{1}}\xi_{j_{2}}^{*})\neq0$, then 
by the continuity of $h'$, we can take a small $\eta \neq 0$ to satisfy the 
following~conditions:~${\rm sgn}~\eta={\rm sgn}~h'(0)$ and 
${\rm sgn}~h'(x)={\rm sgn}~h'(0)$ for all $x$ between $0$ and $\eta$. On the other hand, 
if $h'(0)={\rm sgn}(\alpha_{j_{2}}\xi_{j_{1}}^{*}-\alpha_{j_{1}}\xi_{j_{2}}^{*})=0$, then 
by the continuity of $h''$, we can take a small $\eta \neq 0$ to satisfy~the 
following conditions: ${\rm sgn}~\eta=\pm 1$ and 
${\rm sgn}~h''(x)={\rm sgn}~h''(0)$ for all $x$ between $0$ and $\eta$. 
Then~in either case, it follows from the mean value theorem that 
%
%, and choose $\eta\neq 0$ such that~${\rm sgn}~\eta = 
%{\rm sgn}(\alpha_{j_{2}}\xi_{j_{1}}-\alpha_{j_{1}}\xi_{j_{2}})$, where 
%${\rm sgn~} \eta =\pm 1$  is allowed if 
%$\alpha_{j_{2}}\xi_{j_{1}}-\alpha_{j_{1}}\xi_{j_{2}}=0$. 
%Since~$h'$~and~$h''$~are~continuous, we can take $|\eta| > 0$ 
%small enough so that for all $x$ with $0 < |x| < |\eta|$, 
%%the equality 
%${\rm sgn}~h'( x ) = {\rm sgn}~h'(0)$ holds when   
%${\rm sgn}~h'(0) \neq 0$ and 
%the equality 
%${\rm sgn}~h''( x ) ={\rm sgn}~h''(0)$~holds~when~${\rm sgn}~h'(0) = 0$.~Then %we~have~that 
%
%If ${\rm sgn}(\frac{\xi_{j_{1}}}{\alpha_{j_{1}}}-
%\frac{\xi_{j_{2}}}{\alpha_{j_{2}}})=0$, then from Claim \ref{claim-tech}-(ii), 
%we can arbitrarily take $\eta\neq 0$ such that   
%
\[
\sqrt{\alpha_{j_{1}}^{2}+(\xi_{j_{1}}^{*})^{2}}+\sqrt{\alpha_{j_{2}}^{2}+(\xi_{j_{2}}^{*})^{2}}
< \sqrt{\alpha_{j_{1}}^{2}+(\xi_{j_{1}}^{*}+\eta)^{2}}
+\sqrt{\alpha_{j_{2}}^{2}+(\xi_{j_{2}}^{*}-\eta)^{2}}, 
\]
where the inequality follows from Claim \ref{claim-tech}-(i) for the case that  
$\alpha_{j_{2}}\xi_{j_{1}}^{*}-\alpha_{j_{1}}\xi_{j_{2}}^{*}\neq 0$ 
and~from Claim \ref{claim-tech}-(ii) for the case that 
$\alpha_{j_{2}}\xi_{j_{1}}^{*}-\alpha_{j_{1}}\xi_{j_{2}}^{*}=0$. This implies that 
$G(\xi_{1}^{*},\ldots,\xi_{k}^{*})<G(\xi_{1}',\ldots,\xi_{k}')$, i.e., 
$\vec{\xi}^{*}=(\xi_{1}^{*},\ldots,\xi_{k}^{*}) \in {\cal T}_{f}^{k}$ is not a maximizer of 
$G(\xi_{1},\ldots,\xi_{k})$. 

For the statement (ii), it is immediate from 
the fact that $G(\xi_{1},\ldots,\xi_{k})=G(-\xi_{1},\ldots,-\xi_{k})$~and the definition 
of ${\cal T}_{f}^{k}$, i.e., $-(\xi_{1}+\cdots+\xi_{k})=0$ for any $\vec{\xi} 
=(\xi_{1},\ldots,\xi_{k})\in {\cal T}_{f}^{k}$.  \BQED\medskip%

We have an immediate corollary (to Lemma \ref{lemma-crucial}) that is crucial in the subsequent 
discussions. 
\begin{cor} \label{cor-crucial}
For a maximizer $\vec{\xi}^{*} =(\xi_{1}^{*},\ldots,\xi_{k}^{*}) \in {\cal T}_{f}^{k}$ of the function 
$G(\xi_{1},\ldots,\xi_{k})$,~there~exist
subsets $J_{+},J_{-} \subseteq I=\{1,\ldots,k\}$ that satisfies the following conditions$:$\medskip

{\rm (1)} $J_{+}\cap J_{-}=\emptyset$ and $\leng{J_{+}}+\leng{J_{-}}=k-1;$ 

{\rm (2)} $\xi_{i}^{*} = \beta_{i}$ for each $i \in J_{+};$

{\rm (3)} $\xi_{i}^{*} = -\beta_{i}$ for each $i \in J_{-};$

{\rm (4)} For the index $h \in I\setminus(J_{+}\cup J_{-})$, 
$\sum_{i \in J_{-}} \beta_{i}-\sum_{i \in J_{+}} \beta_{i} = \xi_{h}^{*} \in [-\beta_{h},\beta_{h}]$. 
\end{cor}
From Corollary \ref{cor-crucial}, we have that  
for the maximizer 
$\vec{\xi}^{*} =(\xi_{1}^{*},\ldots,\xi_{k}^{*}) \in {\cal T}_{f}^{k}$ of %the function 
$G(\xi_{1},\ldots,\xi_{k})$, 
\begin{eqnarray}
z_{f}^{k} & = & G(\xi_{1}^{*},\ldots,\xi_{k}^{*})
=\frac{1}{2k}\cdot 
\left\{2\sum_{i=1}^{k}\sqrt{\alpha_{i}^{2}+(\xi_{i}^{*})^{2}}\right\}\nonumber\\
&=&\frac{1}{2k}\cdot \left\{\sqrt{4\alpha_{h}^{2}+(2\xi_{h}^{*})^{2}}+
\sum_{i \in J_{+}}\sqrt{4\alpha_{i}^{2}+(2\xi_{i}^{*})^{2}}+
\sum_{i \in J_{-}}\sqrt{4\alpha_{i}^{2}+(2\xi_{i}^{*})^{2}}\right\}\nonumber\\
& = & \frac{1}{2k}\cdot \left\{\sqrt{4r_{h}+(2\xi_{h}^{*})^{2}}+
\sum_{i \in J_{+}}\sqrt{4r_{i}+(2\beta_{i})^{2}}+
\sum_{i \in J_{-}}\sqrt{4r_{i}+(2\beta_{i})^{2}}\right\}\nonumber\\
& = & \frac{1}{2k}\cdot \left\{\sqrt{4r_{h}+(2\xi_{h}^{*})^{2}}+
\sum_{i \in J_{+}}\sqrt{4r_{i}+(r_{i}-1)^{2}}+
\sum_{i \in J_{-}}\sqrt{4r_{i}+(r_{i}-1)^{2}}\right\}\nonumber\\
& = & \frac{1}{2k}\cdot \left\{\sqrt{4r_{h}+(2\xi_{h}^{*})^{2}}+
\sum_{i \in J_{+}\cup J_{-}} (r_{i}+1)\right\}.\label{eq-crucial}
\end{eqnarray}
If $\leng{\xi_{h}^{*}}=\beta_{h}$, then 
the maximizer 
$\vec{\xi}^{*} =(\xi_{1}^{*},\ldots,\xi_{k}^{*}) \in {\cal T}_{f}^{k}$ of %the function 
$G(\xi_{1},\ldots,\xi_{k})$  
has no unfilled~variables, %i.e., $\leng{\xi_{1}^{*}}=\beta_{1},\ldots,\leng{\xi_{k}^{*}}=\beta_{k}$, 
and we have that 
$4r_{h}+(2\xi_{h}^{*})^{2}=4r_{h}+(2\beta_{h})^{2}=
4r_{h}+(r_{h}-1)^{2}=(r_{h}+1)^{2}$. 
If $\leng{\xi_{h}^{*}}<\beta_{h}$, 
then~the maximizer 
$\vec{\xi}^{*} =(\xi_{1}^{*},\ldots,\xi_{k}^{*}) \in {\cal T}_{f}^{k}$ of %the function 
$G(\xi_{1},\ldots,\xi_{k})$  
has~the~single unfilled variable $\xi_{h}^{*}$ such that 
%$-\beta_{h}<\xi_{h}^{*} <\beta_{h}$ and  
$\sum_{i \in J_{-}}\beta_{i}-\sum_{i\in J_{+}}\beta_{i}=\xi_{h}^{*}\in 
(-\beta_{h},\beta_{h})$, 
and we have that %it follows from Eq.(\ref{eq-crucial}) that 
\[
4r_{h}+(2\xi_{h}^{*})^{2}=4r_{h}+\left(\sum_{i \in J_{-}} 2\beta_{i}-\sum_{i \in J_{+}} 2\beta_{i}\right)^{2}=
4r_{h}+\left\{\sum_{i \in J_{-}} (r_{i}-1)-\sum_{i \in J_{+}} (r_{i}-1)\right\}^{2}.
\]
Thus from Eq.(\ref{eq-crucial}), it follows that for the maximizer 
$\vec{\xi}^{*} =(\xi_{1}^{*},\ldots,\xi_{k}^{*}) \in {\cal T}_{f}^{k}$ of %the function 
$G(\xi_{1},\ldots,\xi_{k})$,~if~the 
%
%\begin{itemize}
%
maximizer $\vec{\xi}^{*} =(\xi_{1}^{*},\ldots,\xi_{k}^{*}) \in {\cal T}_{f}^{k}$ 
of $G$  has no unfilled variable, then 
\begin{equation}
z_{f}^{k}=G(\xi_{1}^{*},\ldots,\xi_{k}^{*}) = \frac{1}{2k}\cdot \sum_{i=1}^{k} (r_{i}+1), \label{eq-crucial-1}
\end{equation}
and if the maximizer $\vec{\xi}^{*} =(\xi_{1}^{*},\ldots,\xi_{k}^{*}) \in {\cal T}_{f}^{k}$ of 
$G$ 
has the single unfilled variable $\xi_{h}^{*}$, then 
\begin{equation}
z_{f}^{k}=G(\xi_{1}^{*},\ldots,\xi_{k}^{*}) = \frac{1}{2k}\cdot \left\{\sqrt{4r_{h}+\left\{
\sum_{i \in J_{-}} (r_{i}-1) - \sum_{i \in J_{+}} (r_{i}-1)\right\}^{2}}+
\sum_{i \in J_{+}\cup J_{-}} (r_{i}+1)\right\}. \label{eq-crucial-2}
\end{equation}
%
%\end{itemize}
%
%=========================================
\subsection{Definitions} \label{subsec-def}
%=========================================
%
%Given $\msc{itv}_{i}=[m_{i},M_{i}]$ for each $1 \leq i \leq k$, 
%we want to maximize the function $H(x_{1},\ldots,x_{k})$~over ${\cal S}_{f}^{k}$. 
From Propositions \ref{prop-correspond} and \ref{prop-max},  it follows that 
maximizing $H(x_{1},\ldots,x_{k})$~over~${\cal S}_{f}^{k}$~is equivalent 
to maximizing $G(\xi_{1},\ldots,\xi_{k})$ over ${\cal T}_{f}^{k}$. 
Our goal is to decide the maximizer $\vec{\xi}^{*}=(\xi_{1}^{*},\ldots,\xi_{k}^{*}) \in 
{\cal T}_{f}^{k}$~of $G(\xi_{1},\ldots,\xi_{k})$, which is equivalent to 
decide $J_{+},J_{-} \subseteq I=\{1,\ldots,k\}$ that satisfies the conditions 
(1), (2), (3), and (4) of Corollary~\ref{cor-crucial}.~We use 
$\msc{amccr}_{k}$ to denote 
the problem of 
deciding~the maximizer 
$\vec{\xi}^{*} =(\xi_{1}^{*},\ldots,\xi_{k}^{*}) \in {\cal T}_{f}^{k}$~of %the function 
$G(\xi_{1},\ldots,\xi_{k})$. 
%\underline{$\msc{amccr}_{k}$}
%
%~~~{\sf Instance:} $\beta_{1}\geq \cdots \beta_{k}$;
%
%~~~{\sf Output:} the maximum of $G(\xi_{1},\ldots,\xi_{k})$ over ${\cal T}_{f}^{k}$.\medskip
%
\begin{df}[($\msc{amccr}_{k}$)] \label{df-maximizer}
For $\beta_{1}\geq \cdots \geq \beta_{k}\geq 0$, find a pair 
$J_{+},J_{-} \subseteq I=\{1,\ldots,k\}$ of index sets that satisfies the conditions 
{\rm (1)}, {\rm (2)}, {\rm (3)}, and {\rm (4)} of {\rm Corollary \ref{cor-crucial}} for a 
maximizer~$\vec{\xi}^{*}=(\xi_{1}^{*},\ldots,\xi_{k}^{*}) \in {\cal T}_{f}^{k}$~of the function 
$G(\xi_{1},\ldots,\xi_{k})$. %: (\xi_{1},\ldots,\xi_{k})\in {\cal T}_{f}^{k}\}$. 
%decide $J_{+},J_{-} \subseteq I=\{1,\ldots,k\}$ that 
%satisfies the conditions of {\rm Corollary \ref{cor-crucial}}. 
%as input, compute    
%the maximizer~$\vec{\xi}^{*}=(\xi_{1}^{*},\ldots,\xi_{k}^{*}) \in 
%{\cal T}_{f}^{k}$ of the function $G(\xi_{1},\ldots,\xi_{k})$. 
%
\end{df}

Given a solution $J_{+},J_{-}$ to $\msc{amccr}_{k}$, 
%maximizer of $G(\xi_{1},\ldots,\xi_{k})$, 
it is easy to write out  
closed formulas of the arithmetic mean component competitive ratio for any integer $k \geq 2$ 
%by Corollary \ref{cor-crucial}. 
as given in the form of Eq.(\ref{eq-crucial-1})~or~Eq.(\ref{eq-crucial-2}). To discuss 
the hardness of $\msc{amccr}_{k}$, we reduce the problem 
{\sc partition} to $\msc{amccr}_{k}$. 
\begin{df}[({\sc partition})] \label{df-partition}
For a set $A=\{a_{1},\ldots,a_{k}\}$ of positive integers, i.e., 
$a_{i} \in {\bf Z}^{+}$~for each $i \in I=\{1,\ldots,k\}$, 
decide whether or not there exists $J \subseteq I$ such that 
%$\sum_{i \in J} a_{i}=\sum_{i \in I\setminus J}a_{i}$
%
\begin{equation}
\sum_{i \in J} a_{i}=
\sum_{i \in I\setminus J}a_{i}. \label{eq-partition}
\end{equation}
\end{df}
It is known that {\sc partition} is ${\cal NP}$-complete \cite{Karp}. 
We say that $A$ is a {\it positive\/}~instance~if~there~exists 
$J \subseteq I$ that satisfies Eq.(\ref{eq-partition}) and  
$A$ is a {\it negative\/} instance if there exists~no~$J \subseteq I$ that satisfies 
Eq.(\ref{eq-partition}). 
Without loss of generality, we assume that~$a_{1}\geq \cdots \geq a_{k}\geq 1$. 
If~$a_{1}+\cdots+a_{k}$~is odd, then there~exist no $J \subseteq I$ that satisfies 
Eq.(\ref{eq-partition}). 
%Then~we~assume that $a_{1}+\cdots+a_{k}$ is even. 
For the case that $a_{1}=\cdots=a_{k}=1$,~there~exists $J \subseteq I$ that satisfies 
Eq.(\ref{eq-partition}) iff $k$ is even. 
Then in the subsequent discussions, 
we assume~that~$a_{1}+\cdots+a_{k}$ is even and there exists $i \in I$~such~that~$a_{i}\geq 2$. 
%
%=========================================================================================
\section{\bm{\mbox{\large {\bf AMCCR}}_{k}} is \bm{{\cal NP}}-Hard} \label{sec-NP-hard}
%=========================================================================================
%
In this section, we show that it is ${\cal NP}$-hard to compute $\msc{amccr}_{k}$ by 
the reduction from 
{\sc partition}~to $\msc{amccr}_{k}$. 
We assume that there exists a polynomial-time 
algorithm~{\sc alg\_a}~for~$\msc{amccr}_{k}$.  
Let $I = \{1,\ldots,k\}$ and $A=\{a_{1},\ldots,a_{k}\}$ be an instance to {\sc partition}. 
%As~mentioned,  assume that $a_{1}\geq 
%\cdots \geq a_{k}\geq 1$, $a_{1}+\cdots+a_{k}$ is even, and 
%there exists $i \in I$~such that $a_{i}\geq 2$. 
Consider the following algorithm {\sc alg\_p} for 
{\sc partition} with access to {\sc alg\_a}~for~$\msc{amccr}_{k}$: \medskip
%reduction from {\sc partition} to $\msc{amccr}_{k}$: \medskip
%
\begin{itembox}[l]{Algorithm: {\sc alg\_p}}
\begin{namelist}{~(3)}
\item[(1)] For the instance $A=\{a_{1},\ldots,a_{k}\}$ to {\sc partition}, set 
$\beta_{i}=a_{i}$ for each $1 \leq i \leq k$; 
\item[(2)] On input $\beta_{1},\ldots,\beta_{k}$, 
run the algorithm {\sc alg\_a} for $\msc{amccr}_{k}$. Let $J_{+},J_{-} \subseteq I=\{1,\ldots,k\}$ 
be the sets returned by {\sc alg\_a} satisfying the conditions 
(1), (2), (3), and (4) of Corollary
\ref{cor-crucial}. A maximizer 
$\vec{\xi}^{*}=(\xi_{1}^{*},\ldots,\xi_{k}^{*}) \in {\cal T}_{f}^{k}$ of 
$G(\xi_{1},\ldots,\xi_{k})$ is induced from $J_{+}$~and~$J_{-}$~according 
to the conditions (2), (3), and (4) of Corollary \ref{cor-crucial}. 
Let $h \in I$ be the index~that~is determined by the condition (4) of Corollary \ref{cor-crucial}. 
%
%
%\hspace*{1.29cm}$\msc{alg\_a}(\beta_{1},\ldots,\beta_{k}) \in {\cal T}_{f}^{k}$  
%be a maximizer of $G(\xi_{1},\ldots,\xi_{k})$ returned by {\sc alg\_a};
%
\item[(3)] Return {\sc yes} if $\leng{\xi_{h}^{*}}=\beta_{h}$; otherwise return {\sc no}. \medskip
\end{namelist}
\end{itembox}\bigskip

%$\vec{\xi}^{*}=(\xi_{1}^{*},\ldots,\xi_{k}^{*})$ has no unfilled variable, 
%and return {\sc no} if $\vec{\xi}^{*}=(\xi_{1}^{*},\ldots,\xi_{k}^{*})$\\
%\hspace*{1.29cm}has the unique unfilled variable $\xi_{h}^{*}$. \medskip 
%
%\begin{equation}
%
%B= \frac{1}{k}\cdot \sum_{i=1}^{k}(\beta_{i}+1). \label{eq-B}
% 
%\end{equation}

\noindent 
On the properties of the above algorithm, the following lemmas hold. 
%
%When the above algorithm returns {\sc yes}, then $A=\{a_{1},\ldots,a_{k}\}$ is indeed a 
%{\it positive\/} instance,~because 
%$\sum_{i \in J_{+}} a_{i} - \sum_{i \in J_{-}} a_{i} \pm a_{h}=0$ holds from the fact that 
%$\vec{\xi}^{*}=(\xi_{1}^{*},\ldots,\xi_{k}^{*}) \in {\cal T}_{f}^{k}$. This implies that 
%either $J=J_{+}\cup \{h\}$ or $J=J_{-}\cup \{h\}$ is a witness of {\sc partition}. On the
%other hand, the above algorithm always returns {\sc no} for a {\it negative\/} instance by the 
%following lemma. 
%
%Then we can show the following lemmas on the properties of $\msc{alg\_a}$ 
%(\beta_{1},\ldots,\beta_{k})$
%for $\msc{amccr}_{k}$.  
%
\begin{lemma} \label{lemma-positive}
If $A=\{a_{1}\ldots,a_{k}\}$ is a {\sf positive} instance, 
then the maximizer~$\vec{\xi}^{*}=(\xi_{1}^{*},\ldots,\xi_{k}^{*}) \in {\cal T}_{f}^{k}$ 
of $G(\xi_{1},\ldots,\xi_{k})$ induced from $(J_{+},J_{-}) =\msc{alg\_a}(\beta_{1},\ldots,\beta_{k})$  
%
%$\vec{\xi}^{*}=(\xi_{1}^{*},\ldots,\xi_{k}^{*})=\msc{alg\_a}(\beta_{1},\ldots,\beta_{k}) \in {\cal T}_{f}^{k}$ 
%of $G(\xi_{1},\ldots,\xi_{k})$ 
has no unfilled~variable.
%
%$\msc{alg\_a}(\beta_{1},\ldots,\beta_{k})=B$. 
%
\end{lemma}
{\bf Proof:} 
If $A=\{a_{1},\ldots,a_{k}\}$ is a positive instance, then there exists $J \subseteq I$ that satisfies 
Eq.(\ref{eq-partition}). By setting $\xi_{i}^{\star}=\beta_{i}=a_{i} \in {\bf Z}$ for each $i \in J$ and 
$\xi_{i}^{\star}=-\beta_{i}=-a_{i} \in {\bf Z}$ for each $i \in I \setminus J$, we have that 
$\vec{\xi}^{\star}=(\xi_{1}^{\star},\ldots,\xi_{k}^{\star}) \in {\cal T}_{f}^{k}$ 
and $\xi_{i}^{\star} \in \msc{itv}_{i}'$ is {\it filled\/} for each $1 \leq i \leq k$. 
Thus %from~the~definition~of $G(\xi_{1},\ldots,\xi_{k})$, 
it follows that 
\[
G(\xi_{1}^{\star},\ldots,\xi_{k}^{\star})=\frac{1}{2k}\sum_{i=1}^{k}
\sqrt{\alpha_{i}^{2}+\beta_{i}^{2}}. 
\]
Let $\vec{\xi}^{*}=(\xi_{1}^{*},\ldots,\xi_{k}^{*}) \in {\cal T}_{f}^{k}$ be a maximizer of 
$G(\xi_{1},\ldots,\xi_{k})$ and 
assume that $\vec{\xi}^{*}
=(\xi_{1}^{*},\ldots,\xi_{k}^{*}) \in {\cal T}_{f}^{k}$ %of $G(\xi_{1},\ldots,\xi_{k})$ 
has an unfilled variable $\xi_{h}$. Then it is immediate that 
\[
0 \leq G(\xi_{1}^{*},\ldots,\xi_{k}^{*})-G(\xi_{1}^{\star},\ldots,\xi_{k}^{\star}) 
= \frac{1}{2k}\left(\sqrt{\alpha_{h}^{2}+(\xi_{h}^{*})^{2}}-
\sqrt{\alpha_{h}^{2}+\beta_{h}^{2}}\right), 
\]
which is possible only if $\xi_{h}^{*}=\pm \beta_{h}$ since 
$\xi_{h}^{*} \in \msc{itv}_{h}'=[-\beta_{h},\beta_{h}]$. This contradicts the assumption that 
$\xi_{h}^{*}$ is an unfilled variable, and complete the proof. \BQED
\begin{lemma} \label{lemma-negative}
If $A=\{a_{1}\ldots,a_{k}\}$ is a {\sf negative} instance, then the maximizer 
$\vec{\xi}^{*}=(\xi_{1}^{*},\ldots,\xi_{k}^{*})\in {\cal T}_{f}^{k}$ 
of $G(\xi_{1},\ldots,\xi_{k})$ induced from $(J_{+},J_{-})=\msc{alg\_a}(\beta_{1},\ldots,\beta_{k})$ 
has a single unfilled variable.
%$\msc{alg\_a}(\beta_{1},\ldots,\beta_{k})\leq B-\frac{1}{2k}$. 
%
\end{lemma}
{\bf Proof:} 
%For a negative instance $A=\{a_{1},\ldots,a_{k}\}$, let 
%$\vec{\xi}^{*}=(\xi_{1}^{*},\ldots,\xi_{k}^{*}) \in {\cal T}_{f}^{k}$ 
%be a maximizer of $G(\xi_{1},\ldots,\xi_{k})$. We show the lemma by contradiction. 
Assume that $\xi_{i}^{*}$ is~filled~for~each~$1 \leq i \leq k$, i.e., 
$\xi_{i}^{*} \in \{-\beta_{i},\beta_{i}\}$ for each $1 \leq i \leq k$,~and~let 
$J=\{i \in I: \xi_{i}^{*}=\beta_{i}\}$. 
Note that $\xi_{1}^{*}+\cdots+\xi_{k}^{*}=0$ for 
$\vec{\xi}^{*}=(\xi_{1}^{*},\ldots,\xi_{k}^{*}) \in {\cal T}_{f}^{k}$. 
%From~the~definition~of ${\cal T}_{f}^{k}$, 
%it is immediate that $\xi_{1}^{*}+\cdots+\xi_{k}^{*}=0$. 
Thus it follows~that 
\[
\sum_{i \in J} a_{i} = \sum_{i \in J} \beta_{i} = 
\sum_{i \in J} \xi_{i}^{*} = -\sum_{i \in I \setminus J} \xi_{i}^{*}
= - \sum_{i \in J} (-\beta_{i})
%- \sum_{i \in I \setminus J} (-\beta_{i})  
= \sum_{i \in I \setminus J} \beta_{i} = \sum_{i \in I \setminus J} a_{i}, 
\]
which implies that $A=\{a_{1},\ldots,a_{k}\}$ is a positive instance. \BQED\medskip

%which contradicts the assumption that $A=\{a_{1},\ldots,a_{k}\}$ 
%is a negative instance. From Corollary \ref{cor-crucial}, it is immediate that 
%the maximizer 
%$\vec{\xi}^{*}=(\xi_{1}^{*},\ldots,\xi_{k}^{*})=\msc{alg\_a}(\beta_{1},\ldots,\beta_{k}) 
%\in {\cal T}_{f}^{k}$~of~the~function 
%$G(\xi_{1},\ldots,\xi_{k})$ has the unique unfilled variable $\xi_{h}^{*}=
%\sum_{i \in J_{-}} \beta_{i}-\sum_{i \in J_{+}} \beta_{i} \in {\bf Z}$. \BQED\medskip
%
%The proofs of Lemmas \ref{lemma-positive} and \ref{lemma-negative} are given in 
%Subsections \ref{proof-positive} and \ref{proof-negative}, respectively. 
Then~from Lemmas 
\ref{lemma-positive} and \ref{lemma-negative}, we can show the following theorem: 
\begin{thm}  \label{thm-NP-hard}
If there exists a polynomial-time algorithm {\sc alg\_a} for $\msc{amccr}_{k}$, 
%with $\ceil{\lg k}+1$ precision after~binary point, 
then~there~exists a polynomial-time algorithm  {\sc alg\_p} for {\sc partition}. 
\end{thm}
{\bf Proof:} Let $A=\{a_{1},\ldots,a_{k}\}$ be an instance to {\sc partition} and 
$\vec{\xi}^{*}=(\xi_{1}^{*},\ldots,\xi_{k}^{*}) \in {\cal T}_{f}^{k}$~be~a~maximizer 
of $G(\xi_{1},\ldots,\xi_{k})$ that is 
induced from $(J_{+},J_{-}) =\msc{alg\_a}(\beta_{1},\ldots,\beta_{k})$. 
If $A$ is a {\it positive\/}~instance, then by Lemma \ref{lemma-positive}, 
the maximizer $\vec{\xi}^{*}=(\xi_{1}^{*},\ldots,\xi_{k}^{*})
\in {\cal T}_{f}^{k}$ of $G(\xi_{1},\ldots,\xi_{k})$ 
has~no~unfilled variable %(Lemma \ref{lemma-positive}), 
and the algorithm {\sc alg\_p} returns {\sc yes} in Step (3). 
If $A$ is a {\it negative\/} instance, then 
by Lemma \ref{lemma-negative}, 
the maximizer $\vec{\xi}^{*}=(\xi_{1}^{*},\ldots,\xi_{k}^{*})\in {\cal T}_{f}^{k}$ 
of $G(\xi_{1},\ldots,\xi_{k})$ has a single unfilled variable %(Lemma \ref{lemma-negative}), 
and the algorithm {\sc alg\_p} returns {\sc no} in Step (3). \BQED\medskip

In Theorem \ref{thm-NP-hard}, we have shown that $\msc{amccr}_{k}$ is ${\cal NP}$-hard
for general integer $k \geq 2$,  
however, 
this does not necessarily imply that 
it is difficult %${\cal NP}$-hard 
to write out closed formulas of the 
arithmetic mean component competitive ratio for a fixed integer $k\geq 2$. 
Then as an application 
of Corollary \ref{cor-crucial}, we derive closed formulas 
of the arithmetic mean component competitive ratio 
for $k=2$, $k=3$, and $k=4$ in Sections \ref{sec-simple}, 
\ref{sec-k=3}, and \ref{sec-k=4}, respectively, in order to enable a precise analysis of the 
multi-objective time series search problem. 
%
%==========================================================
\section{Competitive Ratio for \bm{k=2}} \label{sec-simple}
%==========================================================
%
As we have mentioned, Hasegawa and Itoh \cite{HI} derived a closed formula of the 
arithmetic mean component competitive ratio for $k=2$. In this section, 
we derive a closed formula~of~the~arithmetic mean component competitive 
ratio for $k=2$ given in Eq.(\ref{eq-k=2}) 
as an application of Corollary \ref{cor-crucial} for $k=2$. 
For $\vec{\xi}^{*}=(\xi_{1}^{*},\xi_{2}^{*}) \in {\cal T}_{f}^{2}$ that maximizes 
$G(\xi_{1},\xi_{2})$, let us consider the following~cases: 
(2.0) $\xi_{1}^{*}$ and $\xi_{2}^{*}$ are filled, 
(2.1) $\xi_{1}^{*}$ is unfilled, and 
(2.2) $\xi_{2}^{*}$~is~unfilled. 

The case (2.0) is possible only when $\beta_{1}=\beta_{2}$. From Eq.(\ref{eq-crucial-1}), 
it is immediate that 
\[
G(\pm \beta_{1},\mp\beta_{2}) = %(r_{1}+1)+(r_{2}+1)
G(\pm \beta_{2},\mp\beta_{2})
%= \sqrt{4r_{1}+(r_{1}-1)^{2}}+(r_{2}+1)\\
= \frac{1}{4} \cdot \left\{ \sqrt{4r_{1}+(r_{2}-1)^{2}}+(r_{2}+1)\right\}.
\]
For the case (2.1), we have that $\xi_{2}^{*}$ is filled, i.e., $\xi_{2}^{*}=\pm \beta_{2}$. Since 
$\vec{\xi}^{*}=(\xi_{1}^{*},\xi_{2}^{*}) \in {\cal T}_{f}^{2}$, it is obvious that 
$\xi_{1}=-\xi_{2}=\mp\beta_{2}$. 
Then from Eq.(\ref{eq-crucial-2}), it follows that 
\[
G(\pm \beta_{2},\mp\beta_{2})=\frac{1}{4}\cdot 2\left(
\sqrt{\alpha_{1}^{2}+\beta_{2}^{2}}+\sqrt{\alpha_{2}^{2}+\beta_{2}^{2}} \right)
= \frac{1}{4}\cdot \left\{\sqrt{4r_{1}+(r_{2}-1)^{2}}+(r_{2}+1)\right\}.
\]
For the case (2.2), we have that $\xi_{1}^{*}$ is filled, i.e., $\xi_{1}^{*}=\pm\beta_{1}$. 
Without loss of generality, assume~by Lemma \ref{lemma-crucial}-(ii) that $\xi_{1}^{*}=\beta_{1}$. 
Since $\xi_{2}^{*}$ is unfilled and 
$\vec{\xi}^{*}=(\xi_{1}^{*},\xi_{2}^{*}) \in {\cal T}_{f}^{2}$, 
it is obvious~that~$- \beta_{2}<\xi_{2}^{*}=-\xi_{1}=-\beta_{1}$, which contradicts the 
assumption that $\beta_{1}\geq \beta_{2}\geq 1$. 
%(the same argument holds for $\xi_{1}=-\beta_{1}$). 
Thus the case~(2.2)~never occurs. Then 
the arithmetic mean component competitive ratio $z_{f}^{2}$ is given by 
\[
z_{f}^{2}=G(\pm \beta_{2},\mp\beta_{2})=
\frac{1}{4}\cdot \left\{\sqrt{4r_{1}+(r_{2}-1)^{2}}+(r_{2}+1)\right\}.
\]

In the following sections, we extend the above argument for the case that $k=2$ to the case that 
$k=3$ (see Section \ref{sec-k=3}) and the case that $k=4$ (see Section \ref{sec-k=4}). 
%
%======================================================
\section{Competitive Ratio for \bm{k=3}} \label{sec-k=3}
%======================================================
%
In this section, we derive closed formulas of the arithmetic mean component competitive 
ratio for $k=3$. In fact, we show the following theorems. 
\begin{thm} \label{thm-k=3(1)}
If $(r_{1}-1)\geq (r_{2}-1)+(r_{3}-1)$, then the arithmetic mean component competitive ratio for 
the {\rm 3}-objective time series search problem is 
\[
z_{f}^{3} = \frac{1}{6} \cdot \left[
\sqrt{4r_{1}+\left\{(r_{2}-1)+(r_{3}-1)\right\}^{2}} + (r_{2}+1)+(r_{3}+1)\right]. 
\]
\end{thm}
\begin{thm} \label{thm-k=3(2)}
If $(r_{1}-1)< (r_{2}-1)+(r_{3}-1)$, then the arithmetic mean component competitive ratio for 
the {\rm 3}-objective time series search problem is 
\[
z_{f}^{3} = \frac{1}{6} \cdot \left\{
\sqrt{4r_{3}+(r_{1}-r_{2})^{2}} + (r_{1}+1)+(r_{2}+1)\right\}.
\]
\end{thm}

Let $\vec{\xi}^{*}=(\xi_{1}^{*},\xi_{2}^{*},\xi_{3}^{*}) \in {\cal T}_{f}^{3}$ be a maximizer 
of $G(\xi_{1},\xi_{2},\xi_{3})$.  
By Lemma \ref{lemma-crucial}-(i), 
there~can~exist~a unfilled variable $\xi_{h}^{*} \in \{\xi_{1}^{*},\xi_{2}^{*},\xi_{3}^{*}\}$. 
According to a unfilled variable,  consider the following~cases: 
(3.1) none of the variables $\xi_{1}^{*}$, $\xi_{2}^{*}$, and $\xi_{3}^{*}$ is unfilled or 
the variable $\xi_{1}^{*}$ is unfilled, 
(3.2)~the~variable $\xi_{2}^{*}$ is unfilled, and 
(3.3) the variable $\xi_{3}^{*}$ is unfilled. 

For the case (3.1), two cases (3.1.1) 
$\xi_{1}^{*}=\pm (\beta_{2}+\beta_{3})$ and (3.1.2) $\xi_{1}^{*}=\pm (\beta_{2}-\beta_{3})$ 
are possible, for 
the case (3.2), two cases (3.2.1) 
$\xi_{2}^{*}=\pm (\beta_{1}+\beta_{3})$ and (3.2.2) $\xi_{2}^{*}=\pm (\beta_{1}-\beta_{3})$ 
are possible,~and for the case (3.3), two cases (3.3.1) 
$\xi_{3}^{*}=\pm (\beta_{1}+\beta_{2})$ and (3.3.2) 
$\xi_{3}^{*}=\pm (\beta_{1}-\beta_{2})$~are~possible.~Then 
it is immediate to observe that 
\begin{center}
\begin{tabular}{lcl}
Case (3.1.1) Possible only if $\beta_{1}\geq \beta_{2}+\beta_{3}$. & & 
Case (3.1.2) Always possible.\\
Case (3.2.1) Impossible. & & 
Case (3.2.2) Possible only if $\beta_{1} < \beta_{2}+\beta_{3}$. \\
Case (3.3.1) Impossible. & & 
Case (3.3.2) Possible only if $\beta_{1} < \beta_{2}+\beta_{3}$. 
\end{tabular}
\end{center}
%
%For the case (3.2), %When the variable $\xi_{2}$ is unfilled, 
%we have two cases (3.2.1) 
%$\xi_{2}=\pm (\beta_{1}+\beta_{3})$ and (3.2.2) 
%$\xi_{2}=\pm (\beta_{1}-\beta_{3})$.~Then~it is immediate to observe that 
%
%\begin{center}
%
%\begin{tabular}{rcr}
%
%Case (3.2.1) Impossible. & & 
%Case (3.2.2) Possible only if $\beta_{1} < \beta_{2}+\beta_{3}$. 
%
%\end{tabular}
%
%\end{center}
%
%For the case (3.3), %When the variable $\xi_{3}$ is unfilled, 
%we have two cases (3.3.1) 
%$\xi_{3}=\pm (\beta_{1}+\beta_{2})$ and (3.3.2) 
%$\xi_{3}=\pm (\beta_{1}-\beta_{2})$.~Then~it 
%is immediate to observe that 
%
%\begin{center}
%
%\begin{tabular}{rcr}
%
%Case (3.3.1) Impossible. & & 
%Case (3.3.2) Possible only if $\beta_{1} < \beta_{2}+\beta_{3}$. 
%
%\end{tabular}
%
%\end{center}

We classify the problem instances based on the magnitude of $\beta_{1}$, i.e., 
%we consider the cases 
$\beta_{1}\geq \beta_{2}+\beta_{3}$ and $\beta_{1}< \beta_{2}+\beta_{3}$. 
Then the possibilities for the cases $\mbox{(3.1.1)},\ldots, \mbox{(3.3.2)}$ can be 
summarized in~Table~\ref{fig-k=3}. 
\begin{table}[h!]
\caption{Conditions Among $\beta_{1}$, $\beta_{2}$, and $\beta_{3}$ for $k=3$} 
\label{fig-k=3}
\begin{center}
\def\arraystretch{1.0}
\begin{tabular}{c||c|c} \hline
  &  $\beta_{1} \geq \beta_{2}+\beta_{3}$ & $\beta_{1} < \beta_{2}+\beta_{3}$ \\ \hline\hline
Case (3.1.1) & possible & --- \\ \hline
Case (3.1.2) & possible & possible \\ \hline
Case (3.2.1) & --- & --- \\ \hline
Case (3.2.2) & --- & possible \\ \hline
Case (3.3.1) & --- & --- \\ \hline
Case (3.3.2) & --- & possible \\ \hline
\end{tabular}
\end{center}
\end{table}
%
%\begin{itemize}
%
%\item If $\beta_{1} \geq \beta_{2}+\beta_{3}$, then the cases (3.1.1) and (3.1.2) 
%are possible;
%\item If $\beta_{1} < \beta_{2}+\beta_{3}$, then~the cases 
%(3.1.2), (3.2.2), and (3.3.2) are possible. 
%
%\end{itemize}
%
%Notice that if $\beta_{1} \geq \beta_{2}+\beta_{3}$, then the cases (3.1.1) and (3.1.2) 
%are possible and if $\beta_{1} < \beta_{2}+\beta_{3}$, then~the cases 
%(3.1.2), (3.2.2), and (3.3.2) are possible. 
%
%====================================================================
\subsection{Proof of Theorem \ref{thm-k=3(1)}} \label{proof-k=3(1)}
%====================================================================
%
Assume that $\beta_{1}\geq \beta_{2}+\beta_{3}$, i.e., 
$(r_{1}-1) \geq (r_{2}-1)+(r_{3}-1)$. For the cases (3.1.1)~and~(3.1.2),~we 
use ${\sf V}_{(3.1.1)}$ and ${\sf V}_{(3.1.2)}$ to denote the potential maximum values of 
the function 
$6\cdot G(\xi_{1},\xi_{2},\xi_{3})$~over ${\cal T}_{f}^{3}$, respectively. Then 
from Eq.(\ref{eq-crucial-2}), it follows that 
\begin{eqnarray*}
{\sf V}_{(3.1.1)} & = &6\cdot  G(\pm(\beta_{2}+\beta_{3}),\mp\beta_{2},\mp\beta_{3}) 
= 2\left\{\sqrt{\alpha_{1}^{2}+(\beta_{2}+\beta_{3})^{2}} + 
\sqrt{\alpha_{2}^{2}+\beta_{2}^{2}}+
\sqrt{\alpha_{3}^{2}+\beta_{3}^{2}} \right\}\\
& = & \sqrt{4r_{1}+\{(r_{2}-1)+(r_{3}-1)\}^{2}} + (r_{2}+1)+(r_{3}+1);\\
{\sf V}_{(3.1.2)} & = & 6\cdot G(\pm(\beta_{2}-\beta_{3}),\mp\beta_{2},\pm\beta_{3}) 
= 2 \left\{\sqrt{\alpha_{1}^{2}+(\beta_{2}-\beta_{3})^{2}} + 
\sqrt{\alpha_{2}^{2}+\beta_{2}^{2}}+
\sqrt{\alpha_{3}^{2}+\beta_{3}^{2}} \right\}\\
& = & \sqrt{4r_{1}+\{(r_{2}-1)-(r_{3}-1)\}^{2}} + (r_{2}+1)+(r_{3}+1).
\end{eqnarray*}
%
%\newpage
%
Since $r_{2}\geq r_{3}\geq 1$, it is immediate to see that 
${\sf V}_{(3.1.1)} \geq {\sf V}_{(3.1.2)}$. 
%$G(\pm(\beta_{2}+\beta_{3}),\mp\beta_{2},\mp\beta_{3}) \geq 
%G(\pm(\beta_{2}-\beta_{3}),\mp\beta_{2},\pm\beta_{3})$. 
Thus for the case~that~$(r_{1}-1)\geq (r_{2}-1)+(r_{3}-1)$, we can conclude that 
%
%\begin{eqnarray*}
%
\[
z_{f}^{3} = %\frac{G(\pm(\beta_{2}+\beta_{3}),\mp\beta_{2},\mp\beta_{3})}{6}
\frac{{\sf V}_{(3.1.1)}}{6}
= \frac{1}{6} \cdot \left[\sqrt{4r_{1}+\{(r_{2}-1)+(r_{3}-1)\}^{2}} + (r_{2}+1)+(r_{3}+1)\right].
%
%\end{eqnarray*}
%
\]
%
%=====================================================================
\subsection{Proof of Theorem \ref{thm-k=3(2)}} \label{proof-k=3(2)}
%=====================================================================
%
%To show Theorem \ref{thm-k=3(2)}, 
The following proposition is crucial (and its proof 
is given in Appendix \ref{app-k=3}). 
\begin{prop} \label{prop-k=3}
Let $F_{3}(x,y,z)=\sqrt{4y + (x-z)^2} + (x-y) - \sqrt{4x + (y-z)^2}$. For any 
$x \geq y$, 

{\rm (a)} if $z \leq x$ and $y \geq x-z+1$, then $F_{3}(x,y,z)\geq 0$;

{\rm (b)} if %$z \geq x$ and 
$y \geq z-x+1$, then $F_{3}(x,y,z)\geq 0$. 
\end{prop}

Assume that $\beta_{1}< \beta_{2}+\beta_{3}$, i.e., 
$(r_{1}-1) < (r_{2}-1)+(r_{3}-1)$. For the cases (3.1.2), (3.2.2), and (3.3.2), 
we use ${\sf V}_{(3.1.2)}$, ${\sf V}_{(3.2.2)}$ and ${\sf V}_{(3.3.2)}$ to denote 
the potential maximum 
values~of~the~function $6\cdot G(\xi_{1},\xi_{2},\xi_{3})$, respectively. Then from Eq.(\ref{eq-crucial-2}), 
it follows that 
%
%it is immediate that 
%
\begin{eqnarray}
{\sf V}_{(3.1.2)} & = & 6 \cdot G(\pm(\beta_{2}-\beta_{3}),\mp\beta_{2},\pm\beta_{3}) 
= 2 \left\{
\sqrt{\alpha_{1}^{2}+(\beta_{2}-\beta_{3})^{2}}+ 
\sqrt{\alpha_{2}^{2}+\beta_{2}^{2}}+\sqrt{\alpha_{3}^{2}+\beta_{3}^{2}} \right\}\nonumber\\
& = & \sqrt{4r_{1}+(r_{2}-r_{3})^{2}}+(r_{2}+1)+(r_{3}+1); \label{eq-k=3(1)}\\
{\sf V}_{(3.2.2)} & = & 6 \cdot G(\pm\beta_{1},\mp(\beta_{1}-\beta_{3}),\mp\beta_{3}) 
= 2 \left\{
\sqrt{\alpha_{2}^{2}+(\beta_{1}-\beta_{3})^{2}}+ 
\sqrt{\alpha_{1}^{2}+\beta_{1}^{2}}+\sqrt{\alpha_{3}^{2}+\beta_{3}^{2}} \right\}\nonumber\\
& = & \sqrt{4r_{2}+(r_{1}-r_{3})^{2}}+(r_{1}+1)+(r_{3}+1);\label{eq-k=3(2)}\\
{\sf V}_{(3.3.2)} & = & 6 \cdot G(\pm\beta_{1},\mp\beta_{2},\mp(\beta_{1}-\beta_{2})) 
= 2 \left\{
\sqrt{\alpha_{3}^{2}+(\beta_{1}-\beta_{2})^{2}}+ 
\sqrt{\alpha_{1}^{2}+\beta_{1}^{2}}+\sqrt{\alpha_{2}^{2}+\beta_{2}^{2}} \right\}\nonumber\\
& = & \sqrt{4r_{3}+(r_{1}-r_{2})^{2}}+(r_{1}+1)+(r_{2}+1). \label{eq-k=3(3)}
\end{eqnarray}

In the following lemmas, we show that 
%$G(\pm\beta_{1},\mp\beta_{2},\mp(\beta_{1}-\beta_{2}))$~in~(\ref{eq-k=3(3)})~is~the~maximum 
${\sf V}_{(3.3.2)}$ %by~(\ref{eq-k=3(3)})~
is~the~maximum in ${\sf V}_{(3.1.2)}$, 
${\sf V}_{(3.2.2)}$,~and~${\sf V}_{(3.3.2)}$ for the~case~that $(r_{1}-1) < (r_{2}-1)+(r_{3}-1)$.
\begin{lemma} \label{lemma-k=3(1)}
%
%$G(\pm\beta_{1},\mp(\beta_{1}-\beta_{3}),\mp\beta_{3}) \geq 
%G(\pm\beta_{1},\mp\beta_{2},\mp(\beta_{1}-\beta_{2})) \geq  
%G(\pm(\beta_{2}-\beta_{3}),\mp\beta_{2},\pm\beta_{3})$. 
${\sf V}_{(3.2.2)} \geq {\sf V}_{(3.1.2)}$. 
\end{lemma}
{\bf Proof:} From (\ref{eq-k=3(2)}) and (\ref{eq-k=3(1)}), it is immediate that 
\[
{\sf V}_{(3.2.2)} - {\sf V}_{(3.1.2)}
= \sqrt{4r_{2}+(r_{1}-r_{3})^{2}}+(r_{1}-r_{2})-
\sqrt{4r_{1}+(r_{2}-r_{3})^{2}}. 
\] 
Set $x=r_{1}$, $y=r_{2}$, and $z=r_{3}$. From the fact that $r_{1}\geq r_{2}\geq r_{3}\geq 1$, 
we have that~$x \geq y$~and $z \leq x$, 
and from the assumption that $(r_{1}-1)<(r_{2}-1)+(r_{3}-1)$, it follows that 
$y > x-z+1$. Thus from Proposition \ref{prop-k=3}-(a), the lemma follows. \BQED
%
%where the 1st inequality follows from the assumption~that~$(r_{1}-1)<(r_{2}-1)+(r_{3}-1)$ 
%and~the~2nd inequality follows from the 
%fact that $r_{1}\geq r_{2}\geq r_{3}\geq 1$. \BQED
%
\begin{lemma} \label{lemma-k=3(2)}
%
%$G(\pm\beta_{1},\mp\beta_{2},\mp(\beta_{1}-\beta_{2})) \geq  
%G(\pm\beta_{1},\mp(\beta_{1}-\beta_{3}),\mp\beta_{3})$. 
${\sf V}_{(3.3.2)}\geq {\sf V}_{(3.2.2)}$. 
\end{lemma}
{\bf Proof:} From (\ref{eq-k=3(3)}) and (\ref{eq-k=3(2)}), it is immediate that 
\begin{eqnarray*}
{\sf V}_{(3.3.2)}- {\sf V}_{(3.2.2)} 
& = & 
\sqrt{4r_{3}+(r_{1}-r_{2})^{2}}+(r_{2}-r_{3})- 
\sqrt{4r_{2}+(r_{1}-r_{3})^{2}}\\
& = & 
\sqrt{4r_{3}+(r_{2}-r_{1})^{2}}+(r_{2}-r_{3})- 
\sqrt{4r_{2}+(r_{3}-r_{1})^{2}}. 
\end{eqnarray*}
Set $x=r_{2}$, $y=r_{3}$, and $z=r_{1}$. 
From the fact that $r_{1}\geq r_{2}\geq r_{3}\geq 1$, 
we have that $x \geq y$ %and $z \leq x$, 
and from the assumption that $(r_{1}-1)<(r_{2}-1)+(r_{3}-1)$, it follows that 
$y > z-x+1$. Thus~from Proposition \ref{prop-k=3}-(b), the lemma follows. \BQED\medskip
%
%where the inequality follows from the assumption that $(r_{1}-1)<(r_{2}-1)+(r_{3}-1)$. \BQED\medskip%

From Lemmas \ref{lemma-k=3(1)} and \ref{lemma-k=3(2)}, it follows that 
%$G(\pm\beta_{1},\mp\beta_{2},\mp\beta_{3},
%\pm(-\beta_{1}+\beta_{2}+\beta_{3}))$~
${\sf V}_{(3.3.2)}$ is~the~maximum in ${\sf V}_{(3.1.2)}$, ${\sf V}_{(3.2.2)}$~ and~${\sf V}_{(3.3.2)}$. 
%among the others. 
Thus Theorem \ref{thm-k=3(2)} holds, i.e., if $(r_{1}-1)<(r_{2}-1)+(r_{3}-1)$,~then 
\[
z_{f}^{3} = %\frac{G(\pm\beta_{1},\mp\beta_{2},\mp(\beta_{1}-\beta_{2}))}{6}
\frac{{\sf V}_{(3.3.2)}}{6}
= \frac{1}{6} \cdot \left\{
\sqrt{4r_{3}+(r_{1}-r_{2})^{2}} + (r_{1}+1)+(r_{2}+1)\right\}.
\]
%
%=======================================================
\section{Competitive Ratio for \bm{k=4}} \label{sec-k=4}
%=======================================================
%
In this section, we derive closed formulas of the arithmetic mean component 
competitive~ratio for $k=4$. In fact, we show the following theorems. 
\begin{thm} \label{thm-k=4(1)}
If $(r_{1}-1) \geq (r_{2}-1)+(r_{3}-1)+(r_{4}-1)$, 
then the arithmetic mean component competitive ratio for 
the {\rm 4}-objective time series search problem is 
\[
z_{f}^{4} = \frac{1}{8}\cdot \left[\sqrt{4r_{1}+\left\{(r_{2}-1)+(r_{3}-1)+(r_{4}-1)\right\}^{2}} 
+ (r_{2}+1)+(r_{3}+1)+(r_{4}+1)\right]. 
\]
\end{thm}
\begin{thm} \label{thm-k=4(2)}
If $(r_{2}-1)+(r_{3}-1)-(r_{4}-1) \leq (r_{1}-1) < (r_{2}-1)+(r_{3}-1)+(r_{4}-1)$, 
then the arithmetic mean component competitive ratio for 
the {\rm 4}-objective time series search problem is 
\[
z_{f}^{4} = \frac{1}{8}\cdot \left[\sqrt{4r_{4}+\left\{(r_{1}-1)-(r_{2}-1)-(r_{3}-1)\right\}^{2}} 
+ (r_{1}+1)+(r_{2}+1)+(r_{3}+1)\right].
\]
\end{thm}
\begin{thm} \label{thm-k=4(3)}
If $(r_{1}-1) < (r_{2}-1)+(r_{3}-1)-(r_{4}-1)$, 
then the arithmetic mean component competitive ratio for 
the {\rm 4}-objective time series search problem is 
\[
z_{f}^{4} = \frac{1}{8}\cdot \left[\sqrt{4r_{3}+\left\{(r_{1}-1)-(r_{2}-1)+(r_{4}-1)\right\}^{2}} 
+ (r_{1}+1)+(r_{2}+1)+(r_{4}+1)\right].
\]
\end{thm}

Let $\vec{\xi}^{*}=(\xi_{1}^{*},\xi_{2}^{*},\xi_{3}^{*},\xi_{4}^{*}) \in {\cal T}_{f}^{3}$ be a 
maximizer of $G(\xi_{1},\xi_{2},\xi_{3},\xi_{4})$.  
By Lemma \ref{lemma-crucial}-(i), 
there~can exist a unfilled variable $\xi_{h}^{*} \in \{\xi_{1}^{*},\xi_{2}^{*},\xi_{3}^{*},\xi_{4}^{*}\}$. 
%According to a unfilled variable,  
Let us consider the following cases: 
(4.1) none~of~the 
variables $\xi_{1}^{*}$, $\xi_{2}^{*}$, $\xi_{3}^{*}$, and $\xi_{4}^{*}$ is unfilled or the 
variable $\xi_{1}^{*}$ is unfilled, 
(4.2) the variable $\xi_{2}^{*}$ is unfilled, 
(4.3) the variable $\xi_{3}^{*}$ is unfilled, and 
(4.4) the variable $\xi_{4}^{*}$ is unfilled. 

%When all of the variables $\xi_{1}$, $\xi_{2}$, $\xi_{3}$, and $\xi_{4}$ are filled or the 
%variable $\xi_{1}$ is unfilled,~we~have~four cases 
For the case (4.1), we have 
%For the case that $\xi_{1}$ is unfilled, we have 
four cases 
(4.1.1) $\xi_{1}^{*}=\pm (\beta_{2}+\beta_{3}+\beta_{4})$, 
(4.1.2)~$\xi_{1}^{*}=\pm (\beta_{2}+\beta_{3}-\beta_{4})$, 
(4.1.3) $\xi_{1}^{*}=\pm (\beta_{2}-\beta_{3}+\beta_{4})$, and 
(4.1.4) $\xi_{1}^{*}=\pm (-\beta_{2}+\beta_{3}+\beta_{4})$. 
Then it is immediate~that 
\begin{center}
\begin{tabular}{lcl}
Case (4.1.1) Possible only if $\beta_{1}\geq \beta_{2}+\beta_{3}+\beta_{4}$. & & 
Case (4.1.3) Always possible.\\
Case (4.1.2) Possible only if $\beta_{1}\geq \beta_{2}+\beta_{3}-\beta_{4}$. & & 
Case (4.1.4) Always possible.
\end{tabular}
\end{center}
For the case (4.2), we have four cases 
(4.2.1) $\xi_{2}^{*}=\pm (\beta_{1}+\beta_{3}+\beta_{4})$, 
(4.2.2)~$\xi_{2}^{*}=\pm (\beta_{1}+\beta_{3}-\beta_{4})$, 
(4.2.3) $\xi_{2}^{*}=\pm (\beta_{1}-\beta_{3}+\beta_{4})$, and 
(4.2.4) $\xi_{2}^{*}=\pm (-\beta_{1}+\beta_{3}+\beta_{4})$. % are possible. 
Then it is immediate that 
\begin{center}
\begin{tabular}{lcl}
Case (4.2.1) Impossible. & & 
Case (4.2.3) Possible only if $\beta_{1} < \beta_{2}+\beta_{3}-\beta_{4}$. \\
Case (4.2.2) Impossible. & & 
Case (4.2.4) Possible only if $\beta_{1} < \beta_{2}+\beta_{3}+\beta_{4}$. 
\end{tabular}
\end{center}
For the case (4.3), we have four cases 
(4.3.1) $\xi_{3}^{*}=\pm (\beta_{1}+\beta_{2}+\beta_{4})$,  
(4.3.2)~$\xi_{3}^{*}=\pm (\beta_{1}+\beta_{2}-\beta_{4})$, 
(4.3.3) $\xi_{3}^{*}=\pm (\beta_{1}-\beta_{2}+\beta_{4})$, and 
(4.3.4) $\xi_{3}^{*}=\pm (-\beta_{1}+\beta_{2}+\beta_{4})$. 
Then it is immediate that 
\begin{center}
\begin{tabular}{lcl}
Case (4.3.1) Impossible. & & 
Case (4.3.3) Possible only if $\beta_{1} < \beta_{2}+\beta_{3}-\beta_{4}$. \\
%$\beta_{2}\geq \beta_{1}-\beta_{3}+\beta_{4}$. \\
Case (4.3.2) Impossible. & & 
Case (4.3.4) Possible only if $\beta_{1} < \beta_{2}+\beta_{3}+\beta_{4}$. 
\end{tabular}
\end{center}
For the case (4.4), we have four cases 
(4.4.1) $\xi_{4}^{*}=\pm (\beta_{1}+\beta_{2}+\beta_{3})$, 
(4.4.2)~$\xi_{4}^{*}=\pm (\beta_{1}+\beta_{2}-\beta_{3})$, 
(4.4.3) $\xi_{4}^{*}=\pm (\beta_{1}-\beta_{2}+\beta_{3})$, and 
(4.4.4) $\xi_{4}^{*}=\pm (-\beta_{1}+\beta_{2}+\beta_{3})$. 
Then it is immediate that 
\begin{center}
\begin{tabular}{lcl}
Case (4.4.1) Impossible. & & Case (4.4.3) Impossible.\\
Case (4.4.2) Impossible. & & 
Case (4.4.4) Possible only if $\beta_{2}+\beta_{3}-\beta_{4} < \beta_{1} <
\beta_{2}+\beta_{3}+\beta_{4}$. 
%I
\end{tabular}
\end{center}

We classify the problem instances based on the magnitude of $\beta_{1}$, i.e., 
consider the~cases~$\beta_{1} \geq \beta_{2}+\beta_{3}+\beta_{4}$, 
$\beta_{2} +\beta_{3}-\beta_{4}\leq  \beta_{1}< \beta_{2}+\beta_{3}+\beta_{4}$, and
$\beta_{1} < \beta_{2}+\beta_{3}-\beta_{4}$. 
Then the possibilities~for the cases $\mbox{(4.1.1)},\ldots, \mbox{(4.4.4)}$ can be 
summarized in~Table~\ref{fig-k=4}. 
%
%\begin{itemize}
%
%\item If $\beta_{1} \geq \beta_{2}+\beta_{3}+\beta_{4}$, then the cases 
%(4.1.1), (4.1.2), (4.1.3), and (4.1.4) are possible;
%
%\item If $\beta_{2} +\beta_{3}-\beta_{4}< \beta_{1}< \beta_{2}+\beta_{3}+\beta_{4}$, 
%then the cases (4.1.2), (4.1.3), (4.1.4), (4.2.4),~(4.3.4),~and (4.4.4) are possible; 
%
%\item $\beta_{1} \leq \beta_{2}+\beta_{3}-\beta_{4}$, then the cases 
%(4.1.3), (4.1.4), (4.2.3), (4.2.4), (4.3.3), and (4.3,4) are possible. 
%
%\end{itemize}
%
\begin{table}[h!]
\caption{Conditions Among $\beta_{1}$, $\beta_{2}$, $\beta_{3}$, and 
$\beta_{4}$ for $k=4$} \label{fig-k=4}
\begin{center}
\def\arraystretch{1.0}
\begin{tabular}{c||c|c|c} \hline
  & $\beta_{1} \geq \beta_{2}+\beta_{3}+\beta_{4}$
  & $\beta_{2} +\beta_{3}-\beta_{4} \leq   \beta_{1} <  \beta_{2}+\beta_{3}+\beta_{4}$ 
  & $\beta_{1} < \beta_{2}+\beta_{3}-\beta_{4}$\\ \hline\hline
Case (4.1.1) & possible & --- & --- \\ \hline
Case (4.1.2) & possible & possible & --- \\ \hline
Case (4.1.3) & possible & possible & possible \\ \hline
Case (4.1.4) & possible & possible & possible \\ \hline
Case (4.2.1) & --- & --- & --- \\ \hline
Case (4.2.2) & --- & --- & --- \\ \hline
Case (4.2.3) & --- & --- & possible \\ \hline
Case (4.2.4) & --- & possible & possible \\ \hline
Case (4.3.1) & --- & --- & --- \\ \hline
Case (4.3.2) & --- & --- & --- \\ \hline
Case (4.3.3) & --- & --- & possible \\ \hline
Case (4.3.4) & --- & possible & possible \\ \hline
Case (4.4.1) & --- & --- & --- \\ \hline
Case (4.4.2) & --- & --- & --- \\ \hline
Case (4.4.3) & --- & --- & --- \\ \hline
Case (4.4.4) & --- & possible & --- \\ \hline
\end{tabular}
\end{center}
\end{table}
%
%================================================================
\subsection{Proof of Theorem \ref{thm-k=4(1)}} \label{subsec-r}
%================================================================
%
Assume that $\beta_{1} \geq \beta_{2}+\beta_{3}+\beta_{4}$, i.e., 
$(r_{1}-1) \geq (r_{2}-1)+(r_{3}-1)+(r_{4}-1)$. 
For the cases (4.1.1), (4.1.2), (4.1.3), and (4.1.4), 
let ${\sf V}_{(4.1.1)}$, ${\sf V}_{(4.1.2)}$, ${\sf V}_{(4.1.3)}$, and ${\sf V}_{(4.1.4)}$ 
be the~potential maximum values of the function $8\cdot G(\xi_{1},\xi_{2},\xi_{3},\xi_{4})$, 
respectively. Then From Eq.(\ref{eq-crucial-2}), it follows that 
%
%For the cases~(4.1.1), (4.1.2), (4.1.3), and (4.1.4), it follows that 
%
\begin{eqnarray*}
{\sf V}_{(4.1.1)} & = & 8\cdot G(\pm(\beta_{2}+\beta_{3}+\beta_{4}),\mp\beta_{2},
\mp\beta_{3},\mp\beta_{4})\nonumber\\
& = & 
2 \left\{\sqrt{\alpha_{1}^{2}+(\beta_{2}+\beta_{3}+\beta_{4})^{2}}+ 
\sqrt{\alpha_{2}^{2}+\beta_{2}^{2}} + 
\sqrt{\alpha_{3}^{2}+\beta_{3}^{2}} + 
\sqrt{\alpha_{4}^{2}+\beta_{4}^{2}}\right\}\nonumber\\
& = & \sqrt{4r_{1}+\{(r_{2}-1)+(r_{3}-1)+(r_{4}-1)\}^{2}} + (r_{2}+1)+(r_{3}+1)+(r_{4}+1); \\
%\label{eq-r1}\\
%
{\sf V}_{(4.1.2)} & = & 8\cdot G(\pm(\beta_{2}+\beta_{3}-\beta_{4}),\mp\beta_{2},
\mp\beta_{3},\pm\beta_{4})\nonumber\\
& = & 
2 \left\{\sqrt{\alpha_{1}^{2}+(\beta_{2}+\beta_{3}-\beta_{4})^{2}}+ 
\sqrt{\alpha_{2}^{2}+\beta_{2}^{2}} + 
\sqrt{\alpha_{3}^{2}+\beta_{3}^{2}} + 
\sqrt{\alpha_{4}^{2}+\beta_{4}^{2}}\right\}\nonumber\\
& = & \sqrt{4r_{1}+\{(r_{2}-1)+(r_{3}-1)-(r_{4}-1)\}^{2}} + (r_{2}+1)+(r_{3}+1)+(r_{4}+1); \\
%\label{eq-r2}\\
%
{\sf V}_{(4.1.3)} & = & 8\cdot G(\pm(\beta_{2}-\beta_{3}+\beta_{4}),\mp\beta_{2},
\pm\beta_{3},\mp\beta_{4})\nonumber\\
& = & 
2 \left\{\sqrt{\alpha_{1}^{2}+(\beta_{2}-\beta_{3}+\beta_{4})^{2}}+ 
\sqrt{\alpha_{2}^{2}+\beta_{2}^{2}} + 
\sqrt{\alpha_{3}^{2}+\beta_{3}^{2}} + 
\sqrt{\alpha_{4}^{2}+\beta_{4}^{2}}\right\}\nonumber\\
& = & \sqrt{4r_{1}+\{(r_{2}-1)-(r_{3}-1)+(r_{4}-1)\}^{2}} + (r_{2}+1)+(r_{3}+1)+(r_{4}+1); \\
%\label{eq-r3}\\
%
{\sf V}_{(4.1.4)} & = & 8\cdot G(\pm(-\beta_{2}+\beta_{3}+\beta_{4}),\pm\beta_{2},
\mp\beta_{3},\mp\beta_{4})\nonumber\\
& = & 
2 \left\{\sqrt{\alpha_{1}^{2}+(-\beta_{2}+\beta_{3}+\beta_{4})^{2}}+ 
\sqrt{\alpha_{2}^{2}+\beta_{2}^{2}} + 
\sqrt{\alpha_{3}^{2}+\beta_{3}^{2}} + 
\sqrt{\alpha_{4}^{2}+\beta_{4}^{2}}\right\}\nonumber\\
& = & \sqrt{4r_{1}+\{-(r_{2}-1)+(r_{3}-1)+(r_{4}-1)\}^{2}} + (r_{2}+1)+(r_{3}+1)+(r_{4}+1). 
%\label{eq-r4}
%
\end{eqnarray*}
%
%respectively. 
Since $r_{1}\geq r_{2}\geq r_{3}\geq r_{4}\geq 1$, it is easy to see that 
${\sf V}_{(4.1.1)} \geq {\sf V}_{(4.1.2)}$,~${\sf V}_{(4.1.1)} \geq {\sf V}_{(4.1.3)}$,~and~${\sf V}_{(4.1.1)} \geq {\sf V}_{(4.1.4)}$. 
%
%\begin{eqnarray*}
%
%G(\pm(\beta_{2}+\beta_{3}+\beta_{4}),\mp\beta_{2},
%\mp\beta_{3},\mp\beta_{4}) & \geq & 
%G(\pm(\beta_{2}+\beta_{3}-\beta_{4}),\mp\beta_{2},
%\mp\beta_{3},\pm\beta_{4});\\
%
%G(\pm(\beta_{2}+\beta_{3}+\beta_{4}),\mp\beta_{2},
%\mp\beta_{3},\mp\beta_{4}) & \geq & 
%G(\pm(\beta_{2}-\beta_{3}+\beta_{4}),\mp\beta_{2},
%\pm\beta_{3},\mp\beta_{4});\\
%
%G(\pm(\beta_{2}+\beta_{3}+\beta_{4}),\mp\beta_{2},
%\mp\beta_{3},\mp\beta_{4}) & \geq & 
%G(\pm(-\beta_{2}+\beta_{3}+\beta_{4}),\pm\beta_{2},
%\mp\beta_{3},\mp\beta_{4}), 
%
%\end{eqnarray*}
%
i.e., 
%$G(\pm(\%beta_{2}+\beta_{3}+\beta_{4}),\mp\beta_{2},\mp\beta_{3},\mp\beta_{4})$~
${\sf V}_{(4.1.1)}$ 
is~the maximum %among the others. 
in ${\sf V}_{(4.1.1)}$, ${\sf V}_{(4.1.2)}$, ${\sf V}_{(4.1.3)}$, and ${\sf V}_{(4.1.4)}$. 
Thus for~the~case~that %$\beta_{1}\geq \beta_{2}+\beta_{3}+\beta_{4}$, 
$(r_{1}-1)\geq (r_{2}-1)+(r_{3}-1)+(r_{4}-1)$, 
we can conclude that 
%
%\begin{eqnarray*}
\[
z_{f}^{4} = 
%\frac{G(\pm(\beta_{2}+\beta_{3}+\beta_{4}),\mp\beta_{2},\mp\beta_{3},\mp\beta_{4})}{8}\\
\frac{{\sf V}_{(4.1.1)}}{8}\\
= \frac{1}{8} \cdot \left[
\sqrt{4r_{1}+\{(r_{2}-1)+(r_{3}-1)+(r_{4}-1)\}^{2}} + (r_{2}+1)+(r_{3}+1)+(r_{4}+1)\right].
\]
%
%\end{eqnarray*}
%
%=================================================================
\subsection{Proof of Theorem \ref{thm-k=4(2)}} \label{subsec-m}
%=================================================================
%
%To show Theorem \ref{thm-k=4(2)}, 
The following proposition is crucial (and its proof is given in Appendix 
\ref{app-k=4}). 
\begin{prop} \label{prop-k=4}
Let $F_{4}(x,y,z, p)=\sqrt{4y + (z-x-p)^2} + (x-y) - \sqrt{4x + (z-y-p)^2}$.~For~any $x \geq y$, 
if $y \geq z-x-p+1$, then $F_{4}(x,y,z,p) \geq 0$. 
\end{prop}

Assume that $\beta_{2}+\beta_{3}-\beta_{4}  \leq \beta_{1} 
< \beta_{2}+\beta_{3}+\beta_{4}$, i.e., 
\[
(r_{2}-1)+(r_{3}-1)-(r_{4}-1) \leq (r_{1}-1) < (r_{2}-1)+(r_{3}-1)+(r_{4}-1).
\]
For the cases (4.1.2), (4.1.3), and (4.1.4), 
let ${\sf V}_{(4.1.2)}$, ${\sf V}_{(4.1.3)}$, and ${\sf V}_{(4.1.4)}$ be 
the potential maximum values of the function $8\cdot G(\xi_{1},\xi_{2},\xi_{3},\xi_{4})$, respectively. 
Then from Eq.(\ref{eq-crucial-2}), it follows that 
%
%For the cases (4.1.2), (4.1.3), and (4.1.4), it follows that 
%
\begin{eqnarray}
{\sf V}_{(4.1.2)} & = & 
8\cdot G(\pm(\beta_{2}+\beta_{3}-\beta_{4}),\mp\beta_{2},\mp\beta_{3},
\pm\beta_{4})\nonumber\\
& = & 
2 \left\{\sqrt{\alpha_{1}^{2}+(\beta_{2}+\beta_{3}-\beta_{4})^{2}}+ 
\sqrt{\alpha_{2}^{2}+\beta_{2}^{2}} + 
\sqrt{\alpha_{3}^{2}+\beta_{3}^{2}} + 
\sqrt{\alpha_{4}^{2}+\beta_{4}^{2}}\right\}\nonumber\\
& = & \sqrt{4r_{1}+\{(r_{2}-1)+(r_{3}-1)-(r_{4}-1)\}^{2}} + (r_{2}+1)+(r_{3}+1)+(r_{4}+1); 
\label{eq-m1}\\
{\sf V}_{(4.1.3)} & = & 
8\cdot G(\pm(\beta_{2}-\beta_{3}+\beta_{4}),\mp\beta_{2},\pm\beta_{3},
\mp\beta_{4})\nonumber\\
& = & 
2 \left\{\sqrt{\alpha_{1}^{2}+(\beta_{2}-\beta_{3}+\beta_{4})^{2}}+ 
\sqrt{\alpha_{2}^{2}+\beta_{2}^{2}} + 
\sqrt{\alpha_{3}^{2}+\beta_{3}^{2}} + 
\sqrt{\alpha_{4}^{2}+\beta_{4}^{2}}\right\}\nonumber\\
& = & \sqrt{4r_{1}+\{(r_{2}-1)-(r_{3}-1)+(r_{4}-1)\}^{2}} + (r_{2}+1)+(r_{3}+1)+(r_{4}+1); \nonumber\\
%\label{eq-m2}\\
%
{\sf V}_{(4.1.4)} & = & 
8\cdot G(\pm(-\beta_{2}+\beta_{3}+\beta_{4}),\pm\beta_{2},\mp\beta_{3},
\mp\beta_{4})\nonumber\\
& = & 
2 \left\{\sqrt{\alpha_{1}^{2}+(-\beta_{2}+\beta_{3}+\beta_{4})^{2}}+ 
\sqrt{\alpha_{2}^{2}+\beta_{2}^{2}} + 
\sqrt{\alpha_{3}^{2}+\beta_{3}^{2}} + 
\sqrt{\alpha_{4}^{2}+\beta_{4}^{2}}\right\}\nonumber\\
& = & \sqrt{4r_{1}+\{-(r_{2}-1)+(r_{3}-1)+(r_{4}-1)\}^{2}} + (r_{2}+1)+(r_{3}+1)+(r_{4}+1). \nonumber
%\label{eq-m3}
%
\end{eqnarray}
Since $r_{1}\geq r_{2} \geq r_{3}\geq r_{4}\geq 1$, it is obvious that 
${\sf V}_{(4.1.2)} \geq {\sf V}_{(4.1.3)}$ and ${\sf V}_{(4.1.2)} 
\geq {\sf V}_{(4.1.4)}$,~i.e.,~${\sf V}_{(4.1.2)}$ is the maximum %among the others. 
among ${\sf V}_{(4.1.2)}$, ${\sf V}_{(4.1.3)}$, and ${\sf V}_{(4.1.4)}$. 
For the cases~(4.2.4), (4.3.4), and (4.4.4),~let ${\sf V}_{(4.2.4)}$, ${\sf V}_{(4.3.4)}$, and ${\sf V}_{(4.4.4)}$ be the potential maximum 
values of the function $8\cdot G(\xi_{1},\xi_{2},\xi_{3},\xi_{4})$, respectively. Then 
from Eq.(\ref{eq-crucial-2}), it follows that 
\begin{eqnarray}
{\sf V}_{(4.2.4)} & = & 
8\cdot G(\pm\beta_{1},\pm(-\beta_{1}+\beta_{3}+\beta_{4}),\mp\beta_{3},
\mp\beta_{4})\nonumber\\
& = & 
2 \left\{\sqrt{\alpha_{2}^{2}+(-\beta_{1}+\beta_{3}+\beta_{4})^{2}}+ 
\sqrt{\alpha_{1}^{2}+\beta_{1}^{2}} + 
\sqrt{\alpha_{3}^{2}+\beta_{3}^{2}} + 
\sqrt{\alpha_{4}^{2}+\beta_{4}^{2}}\right\}\nonumber\\
& = & 
2 \left\{\sqrt{\alpha_{2}^{2}+(\beta_{1}-\beta_{3}-\beta_{4})^{2}}+ 
\sqrt{\alpha_{1}^{2}+\beta_{1}^{2}} + 
\sqrt{\alpha_{3}^{2}+\beta_{3}^{2}} + 
\sqrt{\alpha_{4}^{2}+\beta_{4}^{2}}\right\}\nonumber\\
& = & \sqrt{4r_{2}+\{(r_{1}-1)-(r_{3}-1)-(r_{4}-1)\}^{2}} + (r_{1}+1)+(r_{3}+1)+(r_{4}+1); 
\label{eq-m4}\\
{\sf V}_{(4.3.4)} & = & 
8\cdot G(\pm\beta_{1},\mp\beta_{2},\pm(-\beta_{1}+\beta_{2}+\beta_{4}),
\mp\beta_{4})\nonumber\\
& = & 
2 \left\{\sqrt{\alpha_{3}^{2}+(-\beta_{1}+\beta_{2}+\beta_{4})^{2}}+ 
\sqrt{\alpha_{1}^{2}+\beta_{1}^{2}} + 
\sqrt{\alpha_{2}^{2}+\beta_{2}^{2}} + 
\sqrt{\alpha_{4}^{2}+\beta_{4}^{2}}\right\}\nonumber\\
& = & 
2 \left\{\sqrt{\alpha_{3}^{2}+(\beta_{1}-\beta_{2}-\beta_{4})^{2}}+ 
\sqrt{\alpha_{1}^{2}+\beta_{1}^{2}} + 
\sqrt{\alpha_{2}^{2}+\beta_{2}^{2}} + 
\sqrt{\alpha_{4}^{2}+\beta_{4}^{2}}\right\}\nonumber\\
& = & \sqrt{4r_{3}+\{(r_{1}-1)-(r_{2}-1)-(r_{4}-1)\}^{2}} + (r_{1}+1)+(r_{2}+1)+(r_{4}+1); 
\label{eq-m5}\\
{\sf V}_{(4.4.4)} & = & 
8\cdot G(\pm\beta_{1},\mp\beta_{2},\mp\beta_{3},
\pm(-\beta_{1}+\beta_{2}+\beta_{3}))\nonumber\\
& = & 
2 \left\{\sqrt{\alpha_{4}^{2}+(-\beta_{1}+\beta_{2}+\beta_{3})^{2}}+ 
\sqrt{\alpha_{1}^{2}+\beta_{1}^{2}} + 
\sqrt{\alpha_{2}^{2}+\beta_{2}^{2}} + 
\sqrt{\alpha_{3}^{2}+\beta_{3}^{2}}\right\}\nonumber\\
& = & 
2 \left\{\sqrt{\alpha_{4}^{2}+(\beta_{1}-\beta_{2}-\beta_{3})^{2}}+ 
\sqrt{\alpha_{1}^{2}+\beta_{1}^{2}} + 
\sqrt{\alpha_{2}^{2}+\beta_{2}^{2}} + 
\sqrt{\alpha_{3}^{2}+\beta_{3}^{2}}\right\}\nonumber\\
& = & \sqrt{4r_{4}+\{(r_{1}-1)-(r_{2}-1)-(r_{3}-1)\}^{2}} + (r_{1}+1)+(r_{2}+1)+(r_{3}+1). 
\label{eq-m6}
\end{eqnarray} 

In the following lemmas, we show that 
%$G(\pm\beta_{1},\mp\beta_{2},\mp\beta_{3},\pm(-\beta_{1}+\beta_{2}+\beta_{3})$
${\sf V}_{(4.4.4)}$ %in~(\ref{eq-m6})~
is the maximum in ${\sf V}_{(4.1.2)}$, 
${\sf V}_{(4.2.4)}$, ${\sf V}_{(4.3.4)}$,~and ${\sf V}_{(4.4.4)}$ 
for the case that 
$(r_{2}-1)+(r_{3}-1)-(r_{4}-1) \leq (r_{1}-1) < (r_{2}-1)+(r_{3}-1)+(r_{4}-1)$.
%
%$\beta_{2}+\beta_{3}-\beta_{4}  < \beta_{1} 
%< \beta_{2}+\beta_{3}+\beta_{4}$. 
%
\begin{lemma} \label{m6>m1}
%
%$G(\pm\beta_{1},\mp\beta_{2},\mp\beta_{3},
%\pm(-\beta_{1}+\beta_{2}+\beta_{3}))\geq 
%G(\pm(\beta_{2}+\beta_{3}-\beta_{4}),\mp\beta_{2},\mp\beta_{3},
%\pm\beta_{4})$. 
${\sf V}_{(4.4.4)} \geq {\sf V}_{(4.1.2)}$. 
\end{lemma}
{\bf Proof:} From (\ref{eq-m6}) and (\ref{eq-m1}), it is immediate that 
\begin{eqnarray*}
{\sf V}_{(4.4.4)} - {\sf V}_{(4.1.2)}
& = & \sqrt{4r_{4}+(r_{1}-r_{2}-r_{3}+1)^{2}} + (r_{1}-r_{4})-
\sqrt{4r_{1}+(r_{2}+r_{3}-r_{4}-1)^{2}}\\
& = & \sqrt{4r_{4}+(r_{2}+r_{3}-r_{1}-1)^{2}} + (r_{1}-r_{4})-
\sqrt{4r_{1}+(r_{2}+r_{3}-r_{4}-1)^{2}}. 
\end{eqnarray*}
Set $x=r_{1}$, $y=r_{4}$, $z=r_{2}+r_{3}$, and $p=1$. 
From the fact that $r_{1} \geq r_{2} \geq r_{3} \geq r_{4} \geq 1$,~we~have that 
$x\geq y$, and from the 
assumption that $(r_{2}-1)+(r_{3}-1)-(r_{4}-1)\leq  (r_{1}-1)$, it follows~that 
$y \geq  z-x-p+1$. Thus from Proposition \ref{prop-k=4}, the lemma follows. \BQED
\begin{lemma} \label{m6>m4}
${\sf V}_{(4.3.4)} \geq {\sf V}_{(4.2.4)}$. 
\end{lemma}
{\bf Proof:} From (\ref{eq-m5}) and (\ref{eq-m4}), it is immediate that 
\begin{eqnarray*}
{\sf V}_{(4.3.4)} -{\sf V}_{(4.2.4)}
& = & \sqrt{4r_{3}+(r_{1}-r_{2}-r_{4}+1)^{2}} + (r_{2}-r_{3})-
\sqrt{4r_{2}+(r_{1}-r_{3}-r_{4}+1)^{2}}\\
& = & \sqrt{4r_{3}+(r_{1}-r_{4}-r_{2}+1)^{2}} + (r_{2}-r_{3})-
\sqrt{4r_{2}+(r_{1}-r_{4}-r_{3}+1)^{2}}. 
\end{eqnarray*}
Set $x=r_{2}$, $y=r_{3}$, $z=r_{1}-r_{4}$, and $p=-1$. 
From the fact that $r_{1} \geq r_{2} \geq r_{3} \geq r_{4} \geq 1$,~we~have that 
$x\geq y$, and from the 
the assumption~that~$(r_{1}-1)<(r_{2}-1)+(r_{3}-1)+(r_{4}-1)$,~it~follows~that 
$y>z-x-p+1$. Thus from Proposition \ref{prop-k=4}, the lemma follows. \BQED  
\begin{lemma} \label{m6>m5}
${\sf V}_{(4.4.4)} \geq {\sf V}_{(4.3.4)}$. 
\end{lemma}
{\bf Proof:} From (\ref{eq-m6}) and (\ref{eq-m5}), it is immediate that 
\[
{\sf V}_{(4.4.4)} - {\sf V}_{(4.3.4)} 
= \sqrt{4r_{4}+(r_{1}-r_{2}-r_{3}+1)^{2}} + (r_{3}-r_{4})-
\sqrt{4r_{3}+(r_{1}-r_{2}-r_{4}+1)^{2}}. 
%& = & \sqrt{4r_{4}+(-r_{1}+r_{2}+r_{3}-1)^{2}} + (r_{3}-r_{4})-
%\sqrt{4r_{3}+(-r_{1}+r_{2}+r_{4}-1)^{2}}. 
%
\]
Set $x=r_{3}$, $y=r_{4}$, $z=r_{1}-r_{2}$, and $p=-1$. 
From the fact that $r_{1} \geq r_{2} \geq r_{3} \geq r_{4} \geq 1$,~we~have that 
$x\geq y$, and from the 
the assumption that 
$(r_{1}-1)<(r_{2}-1)+(r_{3}-1)+(r_{4}-1)$,~it~follows~that $y>z-x-p+1$. Thus from 
Proposition \ref{prop-k=4}, the lemma follows. \BQED\medskip%

From Lemmas \ref{m6>m1}, \ref{m6>m4}, and \ref{m6>m5}, it is immediate that 
%$G(\pm\beta_{1},\mp\beta_{2},\mp\beta_{3},
%\pm(-\beta_{1}+\beta_{2}+\beta_{3}))$~
${\sf V}_{(4.4.4)}$ is~the~maximum in ${\sf V}_{(4.1.2)}$,~${\sf V}_{(4.2.4)}$, ${\sf V}_{(4.3.4)}$, 
and ${\sf V}_{(4.4.4)}$. 
%among the~others.
Thus Theorem \ref{thm-k=4(2)} holds, i.e., 
%we can conclude that 
if $(r_{2}-1)+(r_{3}-1)-(r_{4}-1) \leq  (r_{1}-1) < (r_{2}-1)+(r_{3}-1)+(r_{4}-1)$, then 
it follows that 
%
%Thus Theorem \ref{thm-k=4(2)} follows 
%from Lemmas \ref{m6>m1}, \ref{m6>m4}, and \ref{m6>m5}, i.e., 
%we can conclude that 
%if $(r_{2}-1)+(r_{3}-1)-(r_{4}-1) < (r_{1}-1) < (r_{2}-1)+(r_{3}-1)+(r_{4}-1)$, then 
%
%\begin{eqnarray*}
\[
z_{f}^{4} =
%\frac{G(\pm\beta_{1},\mp\beta_{2},\mp\beta_{3},\pm(-\beta_{1}+\beta_{2}+\beta_{3}))}{8}\\
\frac{{\sf V}_{(4.4.4)}}{8}
= 
\frac{1}{8}\cdot \left[
\sqrt{4r_{4}+\{(r_{1}-1)-(r_{2}-1)-(r_{3}-1)\}^{2}} + (r_{1}+1)+(r_{2}+1)+(r_{3}+1)\right].
%
%\end{eqnarray*}
\]
%
%===============================================================
\subsection{Proof of Theorem \ref{thm-k=4(3)}} \label{subsec-l}
%===============================================================
%
Assume that $\beta_{1} < \beta_{2}+\beta_{3}-\beta_{4}$, i.e., 
$(r_{1}-1) <  (r_{2}-1)+(r_{3}-1)-(r_{4}-1)$. 
For the cases~(4.1.3), (4.1.4), (4.2.3), (4.2.4), (4.3.3), and (4.3.4), 
we use ${\sf V}_{(4.1.3)}$, ${\sf V}_{(4.1.4)}$, ${\sf V}_{(4.2.3)}$, 
${\sf V}_{(4.2.4)}$, ${\sf V}_{(4.3.3)}$, and ${\sf V}_{(4.3.4)}$ to denote the 
potential maximum values of the function $8\cdot G(\xi_{1},\xi_{2},\xi_{3},\xi_{4})$, respectively. 
Then from Eq.(\ref{eq-crucial-2}), it follows that 
%
%\newpage
%
\begin{eqnarray}
{\sf V}_{(4.1.3)} & = & 
8\cdot G(\pm(\beta_{2}-\beta_{3}+\beta_{4}),\mp\beta_{2},\pm\beta_{3},
\mp\beta_{4})\nonumber\\
& = & 
\sqrt{\alpha_{1}^{2}+(\beta_{2}-\beta_{3}+\beta_{4})^{2}}+ 
\sqrt{\alpha_{2}^{2}+\beta_{2}^{2}} + 
\sqrt{\alpha_{3}^{2}+\beta_{3}^{2}} + 
\sqrt{\alpha_{4}^{2}+\beta_{4}^{2}}\nonumber\\
& = & \sqrt{4r_{1}+\{(r_{2}-1)-(r_{3}-1)+(r_{4}-1)\}^{2}} + (r_{2}+1)+(r_{3}+1)+(r_{4}+1); 
\label{eq-l1}\\
{\sf V}_{(4.1.4)} & = & 
8\cdot G(\pm(-\beta_{2}+\beta_{3}+\beta_{4}),\pm\beta_{2},\mp\beta_{3},
\mp\beta_{4})\nonumber\\
& = & 
\sqrt{\alpha_{1}^{2}+(-\beta_{2}+\beta_{3}+\beta_{4})^{2}}+ 
\sqrt{\alpha_{2}^{2}+\beta_{2}^{2}} + 
\sqrt{\alpha_{3}^{2}+\beta_{3}^{2}} + 
\sqrt{\alpha_{4}^{2}+\beta_{4}^{2}}\nonumber\\
& = & \sqrt{4r_{1}+\{-(r_{2}-1)+(r_{3}-1)+(r_{4}-1)\}^{2}} + (r_{2}+1)+(r_{3}+1)+(r_{4}+1); 
\nonumber\\%\label{eq-l2}\\
{\sf V}_{(4.2.3)} & = & 
8\cdot G(\mp\beta_{1}, \pm(\beta_{1}-\beta_{3}+\beta_{4}),\pm\beta_{3},
\mp\beta_{4})\nonumber\\
& = & 
\sqrt{\alpha_{2}^{2}+(\beta_{1}-\beta_{3}+\beta_{4})^{2}}+ 
\sqrt{\alpha_{1}^{2}+\beta_{1}^{2}} + 
\sqrt{\alpha_{3}^{2}+\beta_{3}^{2}} + 
\sqrt{\alpha_{4}^{2}+\beta_{4}^{2}}\nonumber\\
& = & \sqrt{4r_{2}+\{(r_{1}-1)-(r_{3}-1)+(r_{4}-1)\}^{2}} + (r_{1}+1)+(r_{3}+1)+(r_{4}+1); 
\label{eq-l3}\\
{\sf V}_{(4.2.4)} & = & 
8\cdot G(\pm\beta_{1},\pm(-\beta_{1}+\beta_{3}+\beta_{4}),\mp\beta_{3},
\mp\beta_{4})\nonumber\\
& = & 
\sqrt{\alpha_{2}^{2}+(-\beta_{1}+\beta_{3}+\beta_{4})^{2}}+ 
\sqrt{\alpha_{1}^{2}+\beta_{1}^{2}} + 
\sqrt{\alpha_{3}^{2}+\beta_{3}^{2}} + 
\sqrt{\alpha_{4}^{2}+\beta_{4}^{2}}\nonumber\\
& = & \sqrt{4r_{2}+\{-(r_{1}-1)+(r_{3}-1)+(r_{4}-1)\}^{2}} + (r_{1}+1)+(r_{3}+1)+(r_{4}+1); 
\nonumber\\%\label{eq-l4}\\
{\sf V}_{(4.3.3)} & = & 
8\cdot G(\mp\beta_{1},\pm\beta_{2},\pm(\beta_{1}-\beta_{2}+\beta_{4}),
\mp\beta_{4})\nonumber\\
& = & 
\sqrt{\alpha_{3}^{2}+(\beta_{1}-\beta_{2}+\beta_{4})^{2}}+ 
\sqrt{\alpha_{1}^{2}+\beta_{1}^{2}} + 
\sqrt{\alpha_{2}^{2}+\beta_{2}^{2}} + 
\sqrt{\alpha_{4}^{2}+\beta_{4}^{2}}\nonumber\\
& = & \sqrt{4r_{3}+\{(r_{1}-1)-(r_{2}-1)+(r_{4}-1)\}^{2}} + (r_{1}+1)+(r_{2}+1)+(r_{4}+1); 
\label{eq-l5}\\
{\sf V}_{(4.3.4)} & = & 
8\cdot G(\pm\beta_{1},\mp\beta_{2},\pm(-\beta_{1}+\beta_{2}+\beta_{4}),
\mp\beta_{4})\nonumber\\
& = & 
\sqrt{\alpha_{3}^{2}+(-\beta_{1}+\beta_{2}+\beta_{4})^{2}}+ 
\sqrt{\alpha_{1}^{2}+\beta_{1}^{2}} + 
\sqrt{\alpha_{2}^{2}+\beta_{2}^{2}} + 
\sqrt{\alpha_{4}^{2}+\beta_{4}^{2}}\nonumber\\
& = & \sqrt{4r_{3}+\{-(r_{1}-1)+(r_{2}-1)+(r_{4}-1)\}^{2}} + (r_{1}+1)+(r_{2}+1)+(r_{4}+1). \nonumber
%\label{eq-l6}
%
\end{eqnarray}
From the fact that $r_{1}\geq r_{2}\geq r_{3}\geq r_{4}\geq 1$, 
it is immediate that  
${\sf V}_{(4.1.3)} \geq {\sf V}_{(4.1.4)}$, ${\sf V}_{(4.2.3)} \geq {\sf V}_{(4.2.4)}$, and ${\sf V}_{(4.3.3)} \geq {\sf V}_{(4.3.4)}$. 
%
%\begin{eqnarray*}
%
%G(\pm(\beta_{2}-\beta_{3}+\beta_{4}),\mp\beta_{2},\pm\beta_{3},
%\mp\beta_{4}) & \geq & 
%G(\pm(-\beta_{2}+\beta_{3}+\beta_{4}),\pm\beta_{2},\mp\beta_{3},
%\mp\beta_{4});\\
% 
%G(\mp\beta_{1}, \pm(\beta_{1}-\beta_{3}+\beta_{4}),\pm\beta_{3},
%\mp\beta_{4}) & \geq & 
%G(\pm\beta_{1},\pm(-\beta_{1}+\beta_{3}+\beta_{4}),\mp\beta_{3},
%\mp\beta_{4});\\
%
%G(\mp\beta_{1},\pm\beta_{2},\pm(\beta_{1}-\beta_{2}+\beta_{4}),
%\mp\beta_{4}) & \geq & 
%G(\pm\beta_{1},\mp\beta_{2},\pm(-\beta_{1}+\beta_{2}+\beta_{4}),
%\mp\beta_{4}).
%
%\end{eqnarray*}
%
In the following lemmas, we show that 
%$G(\mp\beta_{1},\pm\beta_{2},\pm(\beta_{1}-\beta_{2}+\beta_{4}),\mp\beta_{4})$
${\sf V}_{(4.3.3)}$ %in~(\ref{eq-l5})~
is~the~maximum in ${\sf V}_{(4.1.3)}$, ${\sf V}_{(4.2.3)}$, and ${\sf V}_{(4.3.3)}$ 
for the case that 
$(r_{1}-1)<  (r_{2}-1)+(r_{3}-1)-(r_{4}-1)$. 
\begin{lemma} \label{l5>l1}
%
%$G(\mp\beta_{1}, \pm(\beta_{1}-\beta_{3}+\beta_{4}),\pm\beta_{3},
%\mp\beta_{4})
%G(\mp\beta_{1},\pm\beta_{2},\pm(\beta_{1}-\beta_{2}+\beta_{4}),\mp\beta_{4})
%\geq 
%G(\pm(\beta_{2}-\beta_{3}+\beta_{4}),\mp\beta_{2},\pm\beta_{3},
%\mp\beta_{4})$.
${\sf V}_{(4.2.3)} \geq {\sf V}_{(4.1.3)}$. 
\end{lemma}
{\bf Proof:} From (\ref{eq-l3}) and (\ref{eq-l1}), it is immediate that 
\begin{eqnarray*} 
{\sf V}_{(4.2.3)} - {\sf V}_{(4.1.3)}
& = & 
\sqrt{4r_{2}+(r_{1}-r_{3}+r_{4}-1)^{2}} + (r_{1}-r_{2})
- \sqrt{4r_{1}+(r_{2}-r_{3}+r_{4}-1)^{2}}\\
& = & 
\sqrt{4r_{2}+(r_{3}-r_{4}-r_{1}+1)^{2}} + (r_{1}-r_{2})
- \sqrt{4r_{1}+(r_{3}-r_{4}-r_{2}+1)^{2}}. 
\end{eqnarray*}
Set $x=r_{1}$, $y=r_{2}$, $z=r_{3}-r_{4}$, and $p=-1$. Since %From the fact that 
$r_{1}\geq r_{2}\geq r_{3}\geq r_{4}\geq 1$, we~have~that~$x\geq y$~and 
\[
y-z+x+p-1= r_{2}-r_{3}+r_{4}+r_{1}-2=(r_{2}-r_{3})+(r_{4}-1)+(r_{1}-1)\geq 0,
\]
i.e., $y\geq z-x-p+1$. 
%from the assumption that $(r_{1}-1)<(r_{2}-1)+(r_{3}-1)-(r_{4}-1)$,~it~follows~that 
%$y>z-x-p+1$. 
Thus from Proposition \ref{prop-k=4}, the lemma follows. \BQED
\begin{lemma} \label{l5>l3}
%
%$G(\mp\beta_{1},\pm\beta_{2},\pm(\beta_{1}-\beta_{2}+\beta_{4}),\mp\beta_{4})
%\geq 
%G(\mp\beta_{1}, \pm(\beta_{1}-\beta_{3}+\beta_{4}),\pm\beta_{3},
%\mp\beta_{4})$. 
${\sf V}_{(4.3.3)} \geq {\sf V}_{(4.2.3)}$. 
\end{lemma}
{\bf Proof:} From (\ref{eq-l5}) and (\ref{eq-l3}), it is immediate that 
\begin{eqnarray*}
{\sf V}_{(4.3.3)} -{\sf V}_{(4.2.3)}
& = & 
\sqrt{4r_{3}+(r_{1}-r_{2}+r_{4}-1)^{2}} + (r_{2}-r_{3}) 
- \sqrt{4r_{2}+(r_{1}-r_{3}+r_{4}-1)^{2}}\\
& = & 
\sqrt{4r_{3}+(r_{1}+r_{4}-r_{2}-1)^{2}} + (r_{2}-r_{3}) 
- \sqrt{4r_{2}+(r_{1}+r_{4}-r_{3}-1)^{2}}. 
\end{eqnarray*}
Set $x=r_{2}$, $y=r_{3}$, $z=r_{1}+r_{4}$, and $p=1$. From the fact that 
$r_{1}\geq r_{2}\geq r_{3}\geq r_{4}\geq  1$, we have that 
$x \geq y$, and 
from the assumption~that~$(r_{1}-1)< (r_{2}-1)+
(r_{3}-1)-(r_{4}-1)$, it follows~that $y> z-x-p+1$. Thus from 
Proposition \ref{prop-k=4}, the lemma follows. \BQED\medskip

From Lemmas \ref{l5>l1} and \ref{l5>l3}, it follows that 
%$G(\mp\beta_{1},\pm\beta_{2},\pm(\beta_{1}-\beta_{2}+\beta_{4}),\mp\beta_{4})$ 
${\sf V}_{(4.3.3)}$ is the maximum in ${\sf V}_{(4.1.3)}$, ${\sf V}_{(4.2.3)}$,~and~${\sf V}_{(4.3.3)}$. 
%among the others. 
Thus Theorem \ref{thm-k=4(3)} holds, i.e., 
if $(r_{1}-1)< (r_{2}-1)+(r_{3}-1)-(r_{4}-1)$, then 
%
%Thus Theorem \ref{thm-k=4(3)} follows from Lemmas \ref{l5>l1} and \ref{l5>l3}, i.e., 
%if $(r_{1}-1)\leq (r_{2}-1)+(r_{3}-1)-(r_{4}-1)$, then 
%
%\begin{eqnarray*}
\[
z_{f}^{4} = 
%\frac{G(\mp\beta_{1},\pm\beta_{2},\pm(\beta_{1}-\beta_{2}+\beta_{4}),\mp\beta_{4})}{8}\\
\frac{{\sf V}_{(4.3.3)}}{8}=
\frac{1}{8}\cdot \left[
\sqrt{4r_{3}+\{(r_{1}-1)-(r_{2}-1)+(r_{4}-1)\}^{2}} + (r_{1}+1)+(r_{2}+1)+(r_{4}+1)\right].
%
%\end{eqnarray*}
\]
%
%===========================

%====================
%
\appendix 
%
%==============================================================
\section{Proof of Claim \ref{claim-tech}} \label{append-claim}
%==============================================================
%
By straightforward calculations, we have that
\begin{eqnarray*}
h'(x) & = & \frac{x+b}{\sqrt{a^{2}+(x+b)^{2}}}+ \frac{x-d}{\sqrt{c^{2}+(d-x)^{2}}};\\
h''(x) & = & \frac{a^{2}}{\{a^{2}+(x+b)^{2}\}^{3/2}}+ \frac{c^{2}}{\{c^{2}+(d-x)^{2}\}^{3/2}}.
\end{eqnarray*}
%
%\newpage
%
\noindent 
It is easy to show that $h'$ and $h''$ are continuous. 
For the statement (ii), it is obvious~that~$h''(0)>0$. For the statement 
(i), it is also immediate that 
\begin{eqnarray*}
{\rm sgn}~h'(0) & = & {\rm sgn} \left(
\frac{b}{\sqrt{a^{2}+b^{2}}} - \frac{d}{\sqrt{c^{2}+d^{2}}}\right)\\
&= &
{\rm sgn} (b \sqrt{c^{2}+d^{2}}-d \sqrt{a^{2}+b^{2}})\\
& = & {\rm sgn} (b^{2}(c^{2}+d^{2})-d^{2}(a^{2}+b^{2}))\\
& =& {\rm sgn}(c^{2}b^{2}-a^{2}d^{2})\\
& =& {\rm sgn}(cb-ad).
%& = & {\rm sgn}\left(
%\frac{b}{a}-\frac{d}{c} \right). 
%
\end{eqnarray*}
%
%=============================
%\section{Proof of Propositions}
%=============================
%
%===============================================================
\section{Proof of Proposition \ref{prop-k=3}} \label{app-k=3}
%===============================================================
%
%Rewrite the function $F_{3}(x,y,z)$ as follows: 
By straightforward calculations, we have that
\begin{eqnarray*}
F_{3}(x,y,z) & = & \sqrt{4y+(x-z)^{2}}+(x-y)-\sqrt{4x+(y-z)^{2}}\\
& = & \frac{\left\{\sqrt{4y+(x-z)^{2}}+(x-y)\right\}^{2}-
\left\{\sqrt{4x+(y-z)^{2}}\right\}^{2}}{\sqrt{4y+(x-z)^{2}}+(x-y)+\sqrt{4x+(y-z)^{2}}}\\
& = & 
\frac{2(x-y)\left\{\sqrt{4y+(x-z)^{2}}+(x-z-2)\right\}}{\sqrt{4y+(x-z)^{2}}+(x-y)+\sqrt{4x+(y-z)^{2}}}.
\end{eqnarray*}

For the statement (a), it is immediate that 
\begin{eqnarray*}
\sqrt{4y+(x-z)^{2}}+(x-z-2) & \geq & \sqrt{4(x-z+1)+(x-z)^{2}}+(x-z-2)\\
& = & \sqrt{(x-z+2)^{2}}+(x-z-2) =2(x-z)\geq 0, 
\end{eqnarray*}
where the 1st inequality follows from the condition that 
$y \geq x-z+1$,  and the last equality~and the 2nd inequality follow from the condition that $z \leq x$. 
Thus the statement (a) holds. 

For the statement (b), it is immediate that 
\begin{eqnarray*}
\sqrt{4y+(x-z)^{2}}+(x-z-2) &\geq & \sqrt{4(z-x+1)+(x-z)^{2}}+(x-z-2)\\
& = & \sqrt{(x-z-2)^{2}}+(x-z-2)\\
&=& \sqrt{(z-x+2)^{2}}-(z-x+2) \\
& = & \leng{z-x+2}-(z-x+2)\\
& \geq &(z-x+2)-(z-x+2)= 0,
\end{eqnarray*}
where the 1st inequality is due to the condition that 
$y \geq z-x+1$. 
% and the last equality follows from the condition that $z \geq x$. 
Thus the statement~(b)~holds. 
%
%===============================================================
\section{Proof of Proposition \ref{prop-k=4}} \label{app-k=4}
%===============================================================
%
Substitute $z-p$ for $z$ in $F_{3}(x,y,z)$ of Proposition \ref{prop-k=3}. Then we have that 
\begin{eqnarray*}
F_{4}(x,y,z,p) & = & F_{3}(x,y,z-p)\\
& = &\sqrt{4y+(x-z+p)^{2}}+(x-y)-\sqrt{4x+(y-z+p)^{2}}\\
& = & \sqrt{4y+(z-x-p)^{2}}+(x-y)-\sqrt{4x+(z-y-p)^{2}}. 
\end{eqnarray*}
Then the proposition immediately follows from Proposition \ref{prop-k=3}-(b). 
%
%
%Rewrite the function $F_{4}(x,y,z,p)$ as follows: 
%
%\begin{eqnarray*}
%
%F_{4}(x,y,z,p) & = & \sqrt{4y + (z-x-p)^2} + (x-y) - \sqrt{4x + (z-y-p)^2}\\
%& = & \frac{\left\{\sqrt{4y + (z-x-p)^2} + (x-y)\right\}^{2} - 
%\left\{\sqrt{4x + (z-y-p)^2}\right\}^{2}}{\sqrt{4y + (z-x-p)^2} 
%+ (x-y) + \sqrt{4x + (z-y-p)^2}}\\
%& = & \frac{2(x-y)\left\{\sqrt{4y+(z-x-p)^{2}}-(z-x-p+2)\right\}}{\sqrt{4y + (z-x-p)^2} 
% + (x-y) + \sqrt{4x + (z-y-p)^2}}\\ 
%& \geq & \frac{2(x-y)\left\{\sqrt{4(z-x-p+1)+(z-x-p)^{2}}
%-(z-x-p+2)\right\}}{\sqrt{4y + (z-x-p)^2} + (x-y) + \sqrt{4x + (z-y-p)^2}}\\
%& = & \frac{2(x-y)\left\{\sqrt{(z-x-p+2)^{2}}
%-(z-x-p+2)\right\}}{\sqrt{4y + (z-x-p)^2} + (x-y) + \sqrt{4x + (z-y-p)^2}} \geq 0
%
%\end{eqnarray*}
%
%where the 1st inequality follows from the condition that $y \geq z-x-p+1$ and the 2nd 
%inequality follows from the assumption that $x \geq y$ and 
%the fact that $\sqrt{a^{2}} = |a|$. 
%

\begin{thebibliography}{99}
%===========================
%
\bibitem{BE} A. Borodin and R. El-Yaniv. Online Computation and 
Competitive Analysis. Cambridge University Press (1998). 
%
\bibitem{E} M. Ehrgott. Multicriteria Optimization. Springer (2005). 
%
\bibitem{Eetal} R. El-Yaniv, A. Fiat, R. M. Karp, and G. Turpin. Optimal Search 
and One-Way Trading Online Algorithms. Algorithmica, 30(1), 
pp.101-139 (2001). 
%
\bibitem{G} M. H. Goldwasser. A Survey of Buffer Management Policies for 
Packet Switches. ACM SIGACT News, 41(1), pp.100-128 (2010). 
%
\bibitem{HI} S. Hasegawa and T. Itoh. 
Optimal Online Algorithms for the Multi-Objective Time Series Search Problem. 
%To appear in {\it Theoretical Computer Science\/} (2017). A preliminary version was appeared 
in Proc. of the 10th International Workshop on Algorithms and Computation, 
WALCOM 2016, Lecture Notes in Computer Science 9627, pp. 201-212 (2016). 
%
\bibitem{Karp} R.M. Karp. Reducibility Among Combinatorial Problems. In Proc. of 
Complexity of Computer Computations, Plenum Press, pp.85-103 (1972). 
%
\bibitem{K} E. Koutsoupias. The $k$-Server Conjecture. 
Computer Science Review, 3(2), pp.105-118 (2009). 
%
\bibitem{MAS} E. Mohr, I. Ahmad, and G. Schmidt. Online Algorithms for 
Conversion Problems: A Survey. Survey in Operations Research 
and Management Science, 19(2), pp.87-104 (2014). 
%
\bibitem{ST} D. D. Sleator and R. Tarjan. Amortized Efficiency of List Update 
and Paging Rules. Communication of the ACM, 28(2), pp.202-208 
(1985). 
%
\bibitem{TIS} M. Tiedemann, J. Ide, and A. Sch\"{o}bel. 
Competitive Analysis for Multi-Objective Online Algorithms. 
in Proc. of the 9th International Workshop on Algorithms and Computation,  
WALCOM 2015, Lecture Notes in Computer Science 8973, pp.210-221 (2015). 
%
\bibitem{Y} N. E. Young. Online Paging and Caching. Encyclopedia of Algorithms, 
Springer (2008). 
%
%====================
\end{thebibliography}
\end{document}